\documentclass{nsr}

\usepackage{amsmath,graphicx,array}
\usepackage{dcolumn,soul}%

\usepackage{amsthm}
\usepackage[figuresright]{rotating}%
\usepackage{algorithm, algorithmicx, algpseudocode}
\usepackage{listings}%
\usepackage{hyperref}

\usepackage{multirow}%
\usepackage{amsfonts}%
\usepackage{mathrsfs}%
\usepackage{xcolor}%
\usepackage{textcomp}%
\usepackage{manyfoot}%
\usepackage{booktabs}%
\usepackage{enumitem}
\usepackage{bm}
\usepackage{diagbox}
\usepackage{physics}
\usepackage{color}
\usepackage{multirow}
\usepackage{makecell}

\makeatletter
\def\uns{\ifmmode\,\else$\,$\fi}%

\newcommand{\mbb}{\mathbb}

\newcommand{\comments}[1]{}

\newcommand{\bs}[1]{\boldsymbol #1}
\newcommand{\EX}{\mathbb{E}}
\newcommand{\be}{\begin{equation}}
\newcommand{\ee}{\end{equation}}
\newcommand{\ds}{\mathrm{ss}}

\newcommand{\xb}{{\boldsymbol x}}

\newcommand{\CDK}{\textrm{CDK}}
\newcommand{\Plk}{\textrm{Plk}}
\newcommand{\APC}{\textrm{APC}}

\renewcommand\eqref[1]{(\ref{#1})}
\newcommand{\myscriptsize}{\fontsize{7pt}{8.4pt}\selectfont}

\definecolor{matplotlib_orange}{RGB}{255, 127, 0}
\definecolor{matplotlib_green}{RGB}{0, 128, 0}

\makeatother
 
\jvol{XX}
\jnum{X}
\jyear{Year}
\doi{10.1093/nsr/XXXX}
\received{XX XX Year}
\revised{XX XX Year}
\accepted{XX XX Year}

\begin{document}

\dhead{RESEARCH ARTICLE}

\subhead{INFORMATION SCIENCE}

\title{EPR-Net: Constructing non-equilibrium potential landscape via a variational force projection formulation}

\author{Yue Zhao$^{1}$}

\author{Wei Zhang$^{2,3,*}$}

\author{Tiejun Li$^{1,4,5,*}$}

\affil{$^1$Center for Data Science, Peking University, Beijing 100871, China}

\affil{$^2$Zuse Institute Berlin, Berlin 14195, Germany}

\affil{$^3$Department of Mathematics and Computer Science, Freie Universit{\"a}t Berlin, Berlin 14195, Germany}

\affil{$^4$LMAM and School of Mathematical Sciences, Peking University, Beijing 100871, China}

\affil{$^5$Center for Machine Learning Research, Peking University, Beijing 100871, China}

\authornote{\textbf{Corresponding authors.} Email: wei.zhang@fu-berlin.de; tieli@pku.edu.cn}

\abstract[ABSTRACT]{We present EPR-Net, a novel and effective deep learning approach that tackles a crucial challenge in biophysics: constructing potential landscapes for high-dimensional non-equilibrium steady-state (NESS) systems. EPR-Net leverages a nice mathematical fact that the desired negative potential gradient is simply the orthogonal projection of the driving force of the underlying dynamics in a weighted inner-product space. Remarkably, our loss function has an intimate connection with the steady entropy production rate (EPR), enabling simultaneous landscape construction and EPR estimation. We introduce an enhanced learning strategy for systems with small noise, and extend our framework to include dimensionality reduction and state-dependent diffusion coefficient case in a unified fashion. Comparative evaluations on benchmark problems demonstrate the superior accuracy, effectiveness, and robustness of EPR-Net compared to existing methods. We apply our approach to challenging biophysical problems, such as an 8D limit cycle and a 52D multi-stability problem, which provide accurate solutions and interesting insights on constructed landscapes. With its versatility and power, EPR-Net offers a promising solution for diverse landscape construction problems in biophysics.}

\maketitle

\section*{INTRODUCTION}\label{sec1}

Since Waddington's famous landscape metaphor on the development of cells in the 1950s \cite{Waddington57}, the construction of potential landscape for non-equilibrium biochemical reaction systems has been recognized as an important problem in theoretical biology, as it provides insightful pictures for understanding complex dynamical mechanisms of biological processes. This problem has attracted considerable attention in recent decades in both biophysics and applied mathematics communities. Until now, different approaches have been proposed to realize Waddington's landscape metaphor in a rational way, and its connection to computing the optimal epigenetic switching paths and switching rates in biochemical reaction systems has also been extensively studied. See~\cite{Ao04,Wang08,Wang10,Wang11,Zhou12,Ge12,Lv14,Shi22,Maier93,Aurell02,Sasai03,Roma05} and the references therein for details and \cite{Ferrell12,Wang15Review,Zhou16,Yuan17,Fang19Review} for reviews. Broadly speaking, these proposals can be classified into two types: (T1) the construction of potential landscape in the finite noise regime \cite{Ao04,Wang08,Wang10,Wang11} and (T2) the construction of the quasi-potential in the zero noise limit \cite{Ge12,Zhou12,Lv14}.

For low-dimensional systems (i.e.,\ dimension less than $4$), the potential landscape can be numerically computed either by solving a Fokker-Planck equation (FPE) using grid-based methods until the steady solution is reached approximately as in (T1) type proposals \cite{Wang08,Li16}, or by solving a Hamilton-Jacobi-Bellman (HJB) equation using, for instance, the ordered upwind method \cite{Cameron12} or minimum action type method \cite{Lv14} as in (T2) type proposals. However, these approaches suffer from the curse of dimensionality when applied to high-dimensional systems. Although methods based on mean-field approximations are able to provide a semi-quantitative description of the potential landscape for some typical systems \cite{Wang10,Li14}, direct and general approaches are still favored in applications. In this aspect, pioneering work has been done recently, which allows direct construction of a high-dimensional potential landscape using deep neural networks (DNNs), based on either the steady viscous HJB equation satisfied by the potential landscape function in the (T1) case~\cite{Lin22,Lin23}, or the \textit{point-wise} orthogonal decomposition of the force field in (T2) case~\cite{Lin21}. 

While these works have significant advanced methodological developments, these approaches are based on solving HJB equations alone and may encounter numerical difficulties due to either non-uniqueness of the weak solution to the non-viscous HJB equation in (T2) case~\cite{Crandall83}, or singularity of the solution with a small noise in (T1) case. Meanwhile, the loss functions considered in \cite{Lin21, Lin22,Lin23} are essentially of physics-informed neural network (PINN) form \cite{Raissi19}, so are generally non-convex, and thus might encounter troubling local minimum issues in the training process.

In this work, we present a simple yet efficient DNN approach to construct the potential landscape of high-dimensional non-equilibrium steady state (NESS) systems in (T1) type with either moderate or small noise. Its intimate connection with non-equilibrium statistical mechanics, nice variational structure and superior numerical performance make it a competitive and promising approach in landscape construction methodology. Our main contributions are as follows.

\begin{enumerate}[label=(\arabic*)]
\item \textit{Proposal of entropy production rate (EPR) loss.} We introduce the convex EPR loss function, reveal its connection with the entropy production rate in statistical physics, and propose its enhanced version using the tempering technique. 
\item \textit{Dimensionality reduction.} We put forward a simple dimensionality reduction strategy when the reducing variables are prescribed, and, interestingly, the reduction formalism has a unified formulation with the EPR framework for the primitive variables. This even holds whenever the system has constant or variable diffusion coefficients.
\item \textit{Successful high-dimensional applications.} We successfully apply our approach to some challenging high-dimensional biological systems including an eight-dimensional (8D) cell cycle model \cite{Wang10} and a 52D multi-stable system \cite{Li13}. Our results reveal more delicate structure of the constructed landscapes than what mean-field approximations typically provide, yet we acknowledge that mean-field approximations are not limited by the potential challenges in SDE simulations.
\end{enumerate}

Overall, EPR-Net offers a promising solution for diverse landscape construction problems in biophysics. Even its nice mathematical structure and connection with non-equilibrium statistical physics make it a unique object that deserves further theoretical and numerical exploration in the future.

\begin{figure*}[h]
\centering
\includegraphics[width=1.0\textwidth]{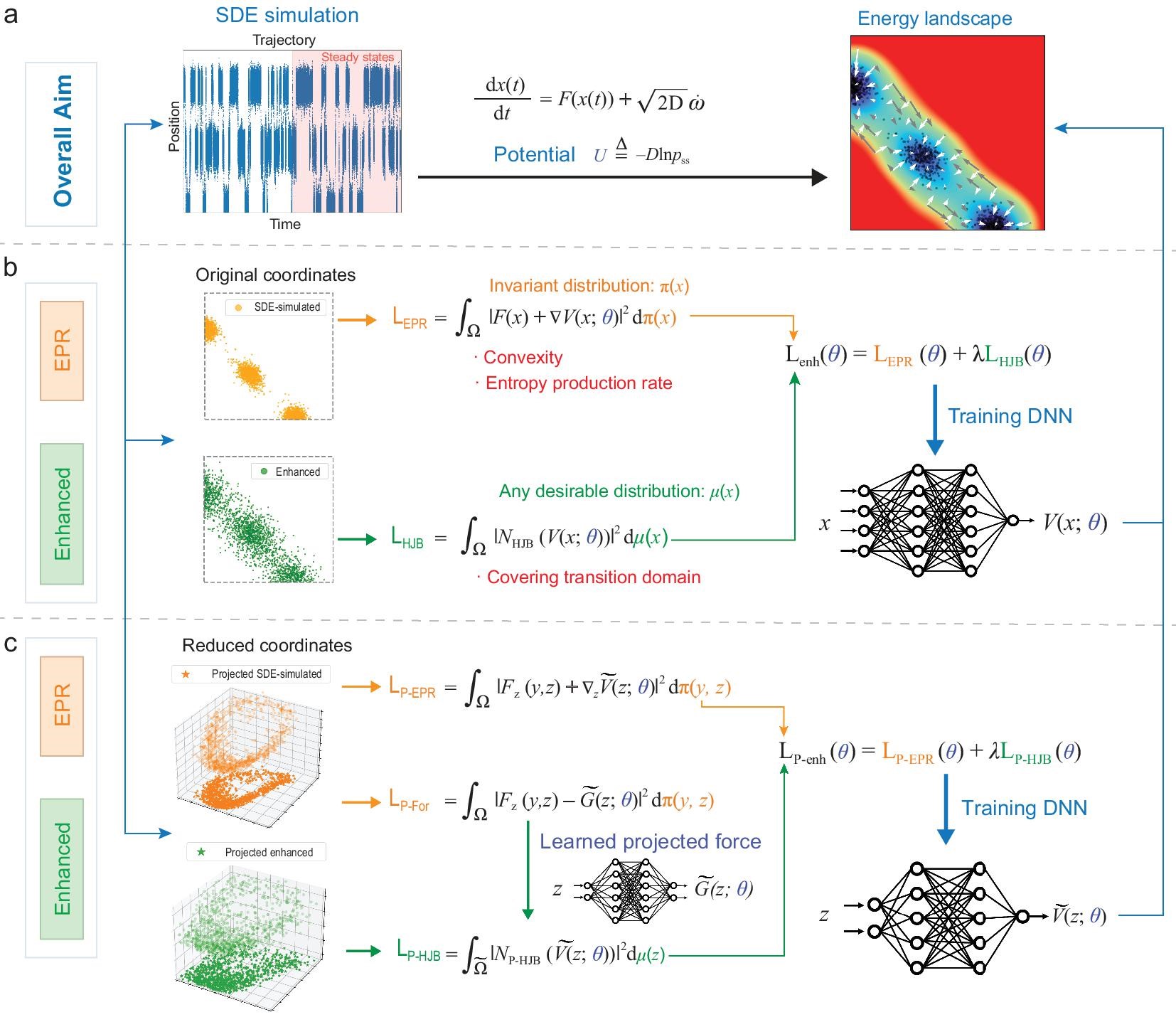}
\caption{\textbf{Constructing energy landscapes through enhanced EPR workflow.} (a) The primary objective is to construct the energy landscape defined through the steady-state distribution of the system. (b) Constructing the high-dimensional energy landscape using the EPR framework with primitive variables. (c) Constructing the dimensionality-reduced energy landscape using EPR with prescribed reduced variables.}
\label{fig:pipeline}
\end{figure*}

\section*{EPR FRAMEWORK}

In this section, we provide an overview of the whole EPR framework, including its primitive formulation, the physical interpretation, its convex property, and the enhanced EPR version.

\subsection*{Overview} \label{sec 2 setup}

Consider an ergodic stochastic differential equation (SDE) 
\be\label{eq:eq1}
\frac{\dd \bs{x}(t)}{\dd t}=\bs{F}(\bs{x}(t))+\sqrt{2D}\,\dot{\bs{w}},\quad \bs{x}(0)=\bs{x}_0,
\ee
where $\bs{x}_0\in \mbb{R}^d$, $\bs{F}: \mathbb{R}^d\rightarrow \mbb{R}^d$ is a smooth driving force,  $\dot{\bs{w}}=(\dot{w}_1,\ldots,\dot{w}_d)^\top$ is the $d$-dimensional temporal Gaussian white noise with $\EX \dot{w}_i(t)=0$ and $\EX [\dot{w}_i(t)\dot{w}_j(s)]=\delta_{ij}\delta(t-s)$ for $i,j=1,\ldots,d$ and $s,t\ge 0$. The constant $D>0$ specifies the noise strength, which is often proportional to the system's temperature $T$. We denote by $p_{\ds}(\bs{x})$ its steady-state probability density function (PDF).

Following the (T1) type proposal in \cite{Wang08}, we define the potential of \eqref{eq:eq1} as $U=-D\ln p_{\ds}$ and the steady probability flux $\bs{J}_{\ds}=p_{\ds}\bs{F}  - D \nabla p_{\ds}$ in the domain $\Omega$, which we assume for simplicity is either $\mbb{R}^d$ or a $d$-dimensional hyperrectangle. The steady-state PDF $p_{\ds}(\bs{x})$ satisfies the Fokker-Planck equation (FPE)
\begin{equation}\label{eq:fp}
-\nabla \cdot(p_{\ds}\bs{F} ) + D \Delta p_{\ds}=0,\quad \textrm{in}~  \Omega,
\end{equation}
and we assume the asymptotic boundary condition (BC) $p_{\ds}(\bs{x})\rightarrow 0$ as $|\bs{x}|\rightarrow \infty$ when $\Omega=\mbb{R}^d$, or the reflecting boundary condition $\bs{J}_{\ds}\cdot \bs{n}=0$ when $\Omega\subset \mbb{R}^d$ is a $d$-dimensional hyperrectangle, where $\bs{n}$ is the unit outer normal. In both cases, we have $p_{\ds}(\bs{x})\ge 0$ and $\int_{\Omega} p_{\ds}(\bs{x})\,\dd\bs{x}=1$.

Here we illustrate the main EPR workflow through Fig.~\textbf{\ref{fig:pipeline}}. As depicted in Fig.~\textbf{\ref{fig:pipeline}a}, our primary objective is to construct the energy landscape $U=-D\ln p_{\ds}$ defined for Eq.~\eqref{eq:eq1}. Leveraging our proposed approach, the enhanced EPR method, we first simulate the considered SDEs until steady state is reached in order to get sample points (the whole `SDE simulation' column of Fig.~\textbf{\ref{fig:pipeline}}). We then train a neural network, representing the potential $U$, for the landscape construction, even when confronted with challenging high-dimensional scenarios. We initially introduce the EPR loss $\operatorname{L_{EPR}}$ (the middle column of Fig.~\textbf{\ref{fig:pipeline}b}), which benefits from its convexity, with its minimum coinciding with the entropy production rate. Subsequently, we present the enhanced EPR loss $\operatorname{L_{enh}}$, tailored to encompass the transition domain more effectively. Furthermore, the overall methodology can be easily extended to the dimensionality reduction problem (Fig.~\textbf{\ref{fig:pipeline}c}), with a unified formulation as the EPR framework shown in Fig.~\textbf{\ref{fig:pipeline}b}, but for the projected variables and corresponding loss functions $\operatorname{L_{P-EPR}}$ and $\operatorname{L_{P-enh}}$.

\subsection*{EPR loss} 

Aiming at an effective approach for high-dimensional applications, we employ NNs to approximate $U(\bs{x})$, and the key idea in this paper is to learn $U$ by training DNN with the loss function
\begin{equation}\label{eq:L-EPR}
\operatorname{L_{EPR}}(V)=\int_{\Omega}|\bs{F}(\bs{x})+\nabla V(\bs{x};\theta)|^2 \,\dd \pi(\bs{x}),
\end{equation}
where $V:=V(\bs{x};\theta)$ is a neural network function with parameters $\theta$ \cite{Goodfellow16}, and $\dd \pi(\bs{x})=p_{\ds}(\bs{x})\, \dd \bs{x}$. To justify \eqref{eq:L-EPR}, 
we note that, for any function $W$, $U$ satisfies the orthogonality relation
\be\label{eq:Orth}
\int_{\Omega} \big(\bs{F}(\bs{x})+\nabla U(\bs{x})\big)\cdot \nabla W(\bs{x})\, \dd \pi(\bs{x}) = 0, 
\ee
which follows from FPE \eqref{eq:fp}, a simple integration by parts and the corresponding BC. Equation \eqref{eq:Orth} means that $\bs{F}+\nabla U$ is perpendicular to the space $\mathcal{G}$ formed by $\nabla W$ for all possible $W$ under the $\pi$-weighted inner product. Equivalently, $-\nabla U$ is the orthogonal projection of the driving force $\bs{F}$ onto $\mathcal{G}$ under the $\pi$-weighted inner product, which implies that $U$ is the unique minimizer (up to a constant) of the loss function \eqref{eq:L-EPR}. See Supplementary Information (SI) Section~\ref{appsec A:validation-EPR} for derivations in detail. We note that related ideas are taken in coarse-graining of molecular systems \cite{deep-cg,cg-md-graph-nn} and generative modeling in machine learning \cite{hyvarinen05a,Song2019}. However, to the best of our knowledge, using the loss in (\ref{eq:L-EPR}) to construct the potential of NESS systems has never been considered before.

In fact, define the $\pi$-weighted inner product for any square integrable functions $f,g$ on $\Omega$ as
\begin{equation}
(f,g)_\pi :=\int_\Omega f(\bs{x})g(\bs{x})\,\dd \pi(\bs{x})
\end{equation}
and the corresponding $L^2_\pi$-norm $\|\cdot\|_\pi$ by $\|f\|^2_{\pi}:= (f,f)_\pi$, we get a Hilbert space $L^2_\pi$ (see, e.g., \cite[Chapter II.1]{Courant53}). Eq.~\eqref{eq:Orth} implies that the minimization of EPR loss finds the orthogonal projection of $\bs{F}$ to the function gradient space $\mathcal{G}$, which is formed by function gradients $\nabla W$ for any $W$,
under the $\pi$-weighted inner product. 
Choosing $W=U$ in \eqref{eq:Orth} gives
\begin{equation}\label{appeq:F-Comp}
\bs{F}(\bs{x}) = -\nabla U(\bs{x})+ \bs{l}(\bs{x}), \text{ such that }  (\nabla U,\bs{l})_\pi = 0.
\end{equation}
However, we remark that this orthogonality holds only in the $L^2_\pi$-inner product sense instead of the pointwise sense.
Furthermore, the two orthogonality relations \eqref{eq:Orth} and \eqref{appeq:F-Comp} can be understood as follows. Using \eqref{appeq:F-Comp}, relation  \eqref{eq:Orth} is equivalent to $\int_{\Omega} \bs{l} \cdot \nabla W d\pi = 0$ for any $W$. Integration by parts gives $\nabla\cdot(\bs{l}\, \mathrm{e}^{-U/D})=0$, which is equivalent to  $ \nabla U \cdot \bs{l}+ D \nabla\cdot\bs{l} = 0$. When $D\rightarrow 0$, we recover the pointwise orthogonality, which is adopted in computing quasi-potentials in \cite{Lin21}. In mathematical language, \eqref{appeq:F-Comp} can be understood as the Hodge-type decomposition in $L^2_\pi$ instead of $L^2$ space.

To utilize (\ref{eq:L-EPR}) in numerical computations, we replace the ensemble average \eqref{eq:L-EPR} with respect to the unknown $\pi$ by the empirical average with data sampled from \eqref{eq:eq1}: 
\begin{equation}\label{eq:L-EPR-T}
\operatorname{\widehat{L}_{EPR}}(\theta)=\frac{1}{N}\sum_{i=1}^N \big|\bs{F}(\bs{x}_i)+\nabla V(\bs{x}_i;\theta)\big|^2.
\end{equation}
Here $(\bs{x}_i)_{1\le i \le N}$ could be either the final states (at time $\mathsf{T}$) of $N$ independent trajectories starting from different initializations, or equally spaced time series along a single long trajectory up to time $\mathsf{T}$, where $\mathsf{T} \gg 1$. In both cases, the ergodicity of SDE in \eqref{eq:eq1} guarantees that (\ref{eq:L-EPR-T}) is a good approximation of (\ref{eq:L-EPR}) as long as $\mathsf{T}$ is large~\cite{Khasminskii12}. We adopt the former approach in the numerical experiments in this work, where
the gradients of both $V$ (with respect to $\bs{x}$) and the loss itself (with respect to $\theta$) in (\ref{eq:L-EPR-T}) are calculated by auto-differentiation through PyTorch~\cite{paszke2019pytorch}. Given the involvement of an approximation on $\pi$, we perform a formal stability analysis of \eqref{eq:L-EPR} in SI Section \ref{Sec B: stability} to ensure its reliability and robustness. Additionally, in the following subsection, we elucidate the benefits of EPR, particularly in terms of \textbf{convexity} and its \textbf{physical interpretation} pertaining to the entropy production rate.

\subsection*{Physical interpretation and convexity}\label{methods: phys}
The minimum loss of \eqref{eq:L-EPR}, denoted as $\operatorname{L_{EPR}}(U)$, possesses a well-defined physical interpretation. 
In the following discussion, we show that this minimum EPR loss aligns precisely with the steady entropy production rate as defined in the NESS theory. Following \cite{Qian01,Zhang12}, we have the important identity concerning the entropy production for~\eqref{eq:eq1}:
\begin{equation}
D\frac{\dd S(t)}{\dd t} = e_p(t)-h_d(t).
\end{equation}
Here $S(t):=-\int_\Omega p(\bs{x},t)\ln p(\bs{x},t)\,\dd \bs{x}$ is the entropy of the probability density function $p(\bs{x},t)$ at time $t$, $e_p$ is the entropy production rate (EPR)
\begin{equation}\label{eqn:epr}
e_p(t) = \int_\Omega \left|\bs{F} - D \nabla \ln p \right|^2 p(\bs{x},t)\, \dd\bs{x},
\end{equation}
and $h_d$ is the heat dissipation rate
\begin{equation}
h_d(t) = \int_\Omega \bs{F}(\bs{x})\cdot \bs{J}(\bs{x},t) \,\dd\bs{x},
\end{equation}
with the probability flux $\bs{J}(\bs{x},t) := p(\bs{x},t)(\bs{F}(\bs{x}) - D\nabla \ln p(\bs{x},t))$ at time $t$.
When $D=k_BT$, the above formulas have clear physical meaning in statistical physics.

From the loss in \eqref{eq:L-EPR} and the steady state of \eqref{eqn:epr}, we get the identity
\begin{equation}\label{eq:L-EPR-U}
\begin{aligned}
\operatorname{L_{EPR}}(U)=&\int_{\Omega} |\bs{J}_{\ds}|^2\frac{1}{p_{\ds}}\, \dd \bs{x}\\
 =&\int_\Omega \left|\bs{F} - D \nabla \ln p_{\ds} \right|^2 p_\ds \,\dd\bs{x}=e^{\rm{ss}}_p,
\end{aligned}
\end{equation}
where $\bs{J}_{\ds}(\bs{x})$ is the steady probability flux and $e^{\rm{ss}}_p$ denotes the steady entropy production rate (EPR) of the NESS system \eqref{eq:eq1} \cite{Qian01,Wang08,Zhang12}. 
Therefore, minimizing \eqref{eq:L-EPR} is equivalent to approximating the steady \textbf{E}ntropy \textbf{P}roduction \textbf{R}ate. This correspondence provides a rationale for naming our approach `EPR-Net'.

EPR loss has an appealing property that it is strictly convex on $V$ (up to a constant), i.e.,
\be\label{eq:convex}
\operatorname{L_{EPR}}(V_\omega)< (1-\omega)\operatorname{L_{EPR}}(V_0)+\omega\operatorname{L_{EPR}}(V_1)
\ee
for any $0<\omega<1$ and $V_0,V_1$ which satisfy $\nabla (V_0-V_1)\not\equiv 0$, where $V_\omega:=(1-\omega) V_0+\omega V_1$. Eq.~\eqref{eq:convex} can be easily verified by direct calculations. This strict convexity guarantees the uniqueness of the critical point $V$ (up to a constant)
and provides theoretical guarantee on the fast convergence of the training procedure under certain assumptions \cite{Mei2018}. It may also contribute to better convergence behavior of EPR during the training process, as mentioned in the numerical comparisons part.

\subsection*{Enhanced EPR loss\label{mt: enhanced epr}} Substituting the relation $p_{\ds}(\bs{x})= \exp(-U(\bs{x})/D)$ into \eqref{eq:fp}, we get the viscous HJB equation
\begin{equation}\label{eq:HJB}
\mathcal{N}_{\textrm{HJB}}(U):= -\bs{F} \cdot \nabla U+ D \Delta U -|\nabla U|^2+D \nabla \cdot \bs{F} =0
\end{equation}
with the asymptotic BC $U\rightarrow \infty$ as $|\bs{x}|\rightarrow \infty$ in the case of $\Omega=\mathbb{R}^d$, or the reflecting BC $(\bs{F}+\nabla U) \cdot \bs{n}=0$ on $\partial \Omega$ when $\Omega$ is a $d$-dimensional hyperrectangle, respectively. As in the framework of PINNs~\cite{Raissi19}, \eqref{eq:HJB} motivates the HJB loss  
\begin{equation}\label{eq:HJB-loss}
\operatorname{L_{HJB}}(V)= \int_{\Omega}\big|\mathcal{N}_{\textrm{HJB}}(V(\bs{x};\theta)) \big|^2 \,\dd \mu(\bs{x}),
\end{equation}
where $\mu$ is any desirable distribution. 
By choosing $\mu$ properly, this loss allows the use of sample data that better covers the domain $\Omega$ and, when combined with the loss in \eqref{eq:L-EPR}, leads to improvement of the training results 
when $D$ is small. Specifically, for small $D$, we propose the enhanced loss which has the form 
\be\label{eq:enh}
\operatorname{\widehat{L}_{enh}}(\theta)=\operatorname{\widehat{L}_{EPR}}(\theta)+\lambda \operatorname{\widehat{L}_{HJB}}(\theta),
\ee
where $\operatorname{\widehat{L}_{EPR}}(\theta)$ is defined in \eqref{eq:L-EPR-T},
$\operatorname{\widehat{L}_{HJB}}(\theta) = \frac{1}{N'}\sum_{i=1}^{N'} |\mathcal{N}_{\textrm{HJB}}(V(\bs{x}'_i;\theta))|^2$ is an approximation of \eqref{eq:HJB-loss} using an independent data set  $(\bs{x}'_i)_{1\le i\le N'}$ obtained by sampling the trajectories of \eqref{eq:eq1} with a larger $D'> D$. The weight parameter $\lambda>0$ balances the contribution of the two terms in \eqref{eq:enh}. While its value can be tuned based on system's properties, in our numerical experiment we observed that the method performs well for $\lambda$ taking values in a relatively broad range. Note that the proposed strategy is both general and easily adaptable. For instance, one can alternatively use data $(\bs{x}'_i)_{1\le i\le N'}$ that contains more samples in the transition region, or employ in \eqref{eq:enh} a modification of the loss \eqref{eq:HJB-loss}~\cite{Lin22}. We further illustrate the motivation of enhanced EPR in SI Section \ref{method: motivation for enhanced epr}.

\section*{RESULTS FOR ENERGY LANDSCAPE CONSTRUCTION}

In this section, we demonstrate the superiority of the enhanced EPR over the HJB loss alone and over the normalizing flow (NF), which is a class of generative models used for density estimation that leverage invertible transformations to map between complex data distributions and simple latent distributions~\cite{Koby21}, through 2D benchmark examples. We then apply the enhanced EPR to the 3D Lorenz model and a 12D Gaussian mixture model to show its effectiveness in constructing energy landscapes in higher dimensions. We remark that alternative approaches have been investigated for potential construction in limit cycles~\cite{Zhu2006Limit} and the Lorenz system~\cite{Ma2014Potential} distinct from our methodology.

\subsection*{Two-dimensional benchmark examples} \label{compare}

Within this subsection, we undertake a comparative analysis of 2D benchmark problems. These encompass a toy model, a 2D biological system exhibiting a limit cycle~\cite{Wang08}, and a 2D multi-stable system~\cite{Wang11}, see SI Section~\ref{Sec F.1.}-\ref{sec:F.4} for training details, additional results and problem settings.

\begin{figure*}[!h]
\begin{center}
\includegraphics[width=1.0\textwidth]{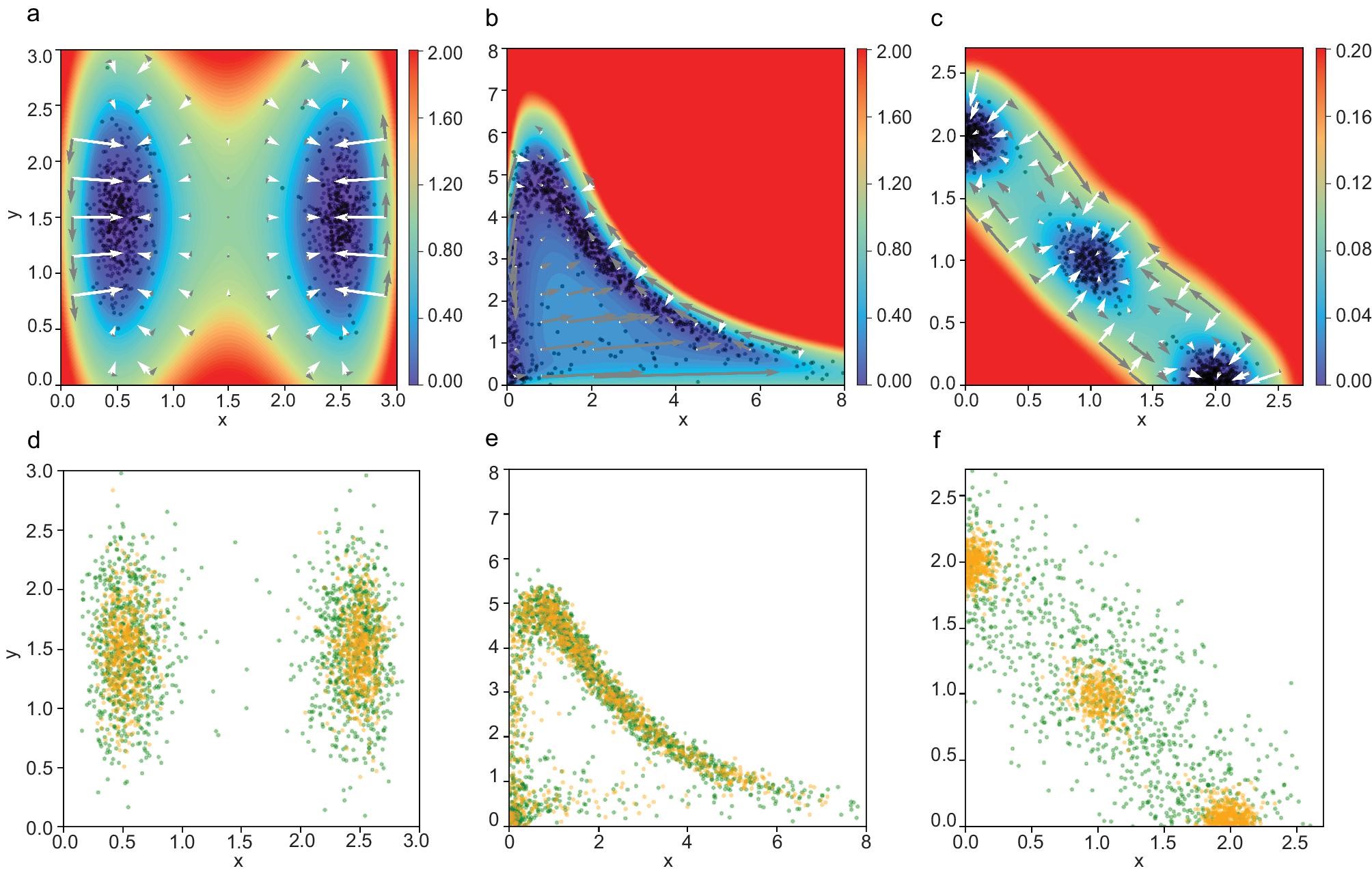}
    \caption{\textbf{Two-dimensional benchmark examples solved under the EPR framework.} (a-c) Filled contour plots of the learned potential $V(\bs{x};\theta^*)$ for (a) a toy model learned by the EPR loss \eqref{eq:L-EPR}, (b) a biochemical oscillation network model \cite{Wang08} and (c) a tri-stable cell development model \cite{Wang11}, all of which are learned by the enhanced loss \eqref{eq:enh}. The force field $\bs{F}(\bs{x})$ is decomposed into the gradient part $-\nabla V(\bs{x};\theta^*)$ (white arrows) and the non-gradient part $\bs{F}(\bs{x})+\nabla V(\bs{x};\theta^*)$ (gray arrows). The length of an arrow corresponds to the magnitude of the vector. The solid dots are samples from the simulated invariant distribution. (d-f) SDE-simulated samples $(\bs{x}_i)_{1\le i\le N}$ (yellow points) and enhanced samples $(\bs{x}'_i)_{1\le i\le N'}$ (green points). (d) $D=0.1$, where enhanced samples are generated with $D'=2D$. (e) $D=0.1$, where enhanced samples are obtained by adding Gaussian perturbations with $\sigma=0.05$ on SDE-simulated samples. (f) $D=0.01$, where enhanced samples are generated with $D'=10D$.}
\label{fig:dw2d_force}
\end{center}
\end{figure*}

\subsubsection*{Constructing landscapes}

The potential function $V(\bs{x}; \theta)$ is approximated using a feedforward neural network architecture consisting of three hidden layers, employing the tanh activation function. Each hidden layer comprises 20 hidden states. Subsequently, we refer to the enhanced loss as
\begin{equation}
\operatorname{L_\text{enh}} = \lambda_1 \operatorname{L_\text{EPR}} + \lambda_2 \operatorname{L_\text{HJB}},
\label{eq:enh-loss-sm}
\end{equation}
where $\lambda_1$ and $\lambda_2$ should be chosen to balance the corresponding loss terms. In SI Section~\ref{Sec F.1.}, we conduct a sensitivity analysis of $\lambda_1$, demonstrating that our methods are robust and effective across a broad spectrum of scenarios.

As illustrated in Fig.~\ref{fig:dw2d_force}a-c, we initially showcase well-constructed landscapes under the EPR framework. These include the acquired potentials $V(\bs{x};\theta^*)$, the decomposition of forces, and sample points derived from the simulated invariant distribution. In the toy model (Fig.~{\ref{fig:dw2d_force}a}), the gradient of the potential (white arrows) points directly towards the corresponding attractor, while the non-gradient part of the force field (gray arrows) shows a counter-clockwise rotation in the model with a limit cycle (Fig.~{\ref{fig:dw2d_force}b}), and a splitting-and-back flow from the attractor in the middle to the other two attractors in the tri-stable dynamical model (Fig.~{\ref{fig:dw2d_force}c}).

Data enhancement techniques employed to generate improved samples are flexible, as visually depicted in Fig.~{\ref{fig:dw2d_force}d-f}. We present choices for obtaining these samples, including generating more diffusive samples with a diffusion coefficient $D' > D$ or introducing Gaussian perturbations with a standard deviation of $\sigma$. We provide a more comprehensive analysis of the data enhancement in SI Section \ref{Sec F.1.},  demonstrating the resilience of enhanced EPR to variations in the enhanced samples. Specifically in Fig. {\ref{fig:dw2d_force}}, we use $D' = 2D$ for the toy model, $D' = 10D$ for the multi-stable problem and $\sigma=0.05$ for the limit-cycle problem. We use the same size of the SDE-simulated dataset $(\bs{x}_i)_{1\le i\le N}$ and the enhanced dataset $(\bs{x}'_i)_{1\le i\le N'}$, denoted as $N = N'$.

\subsubsection*{Numerical comparisons}

We proceed to perform a quantitative and comprehensive comparison of the performance between enhanced EPR, solving HJB alone and NF. For the toy model, the true solution is analytically known (see SI Section \ref{methods: 2d toy model}), while for the other two 2D examples, we compute the reference solution by discretizing the steady FPE using a piecewise bilinear finite element method on a fine rectangular grid and computing the obtained sparse linear system using 
the least squares solver (the normalization condition $\int_{\Omega} p_{\rm{ss}}(\bs{x})\dd \bs{x}=1$ is utilized to fix the additive constant). After training, we shift the minimum of the potentials to the origin and
we measure their accuracy using the relative root mean square error (rRMSE) and the relative mean absolute error (rMAE):
\begin{align}\label{appeq:rrmse}
\operatorname{rRMSE} & =\sqrt{\frac{\int_{\mathcal{D}}\left|V(\bs{x};\theta^{*})-U_0(\bs{x})\right|^2 \dd \bs{x}}{\int_{\mathcal{D}}|U_0(\bs{x})|^2 \dd \bs{x}}}, \\
\operatorname{rMAE} & =\frac{\int_{\mathcal{D}}\left|V(\bs{x};\theta^*)-U_0(\bs{x})\right| \dd \bs{x}}{\int_{\mathcal{D}}|U_0(\bs{x})| \dd \bs{x}},\label{appeq:rmae}
\end{align}
on the domain $\mathcal{D} = \{\bs{x} \in \Omega| U_0(\bs{x}) \leq 20 D\}$,
where $U_0$ denotes the reference solution.

\begin{table*}[!h]
\centering
\caption{Comparisons of different methods. We report the mean and the standard deviation over 5 random seeds. The best results are highlighted in bold.}
\label{tab:compare}
\begin{tabular}{cccc}
 \toprule
\textbf{\textbf{Problem}} & \textbf{\textbf{Method}} & \textbf{\textbf{rRMSE}} & \textbf{\textbf{rMAE}} \\ \midrule
\multirow{3}{*}{Toy, D=0.1}                & Enhanced EPR                              & \textbf{0.028}$\pm${\scriptsize 0.010}        & \textbf{0.026}$\pm${\scriptsize 0.010}       \\
                                           & HJB loss alone                                 &  {0.034}$\pm${\scriptsize 0.016}                                    &  {0.029}$\pm${\scriptsize 0.014}                                   \\
                                           & Normalizing Flow                          &  {0.124}$\pm${\scriptsize 0.016}                                   &  {0.082}$\pm${\scriptsize 0.011}                                   \\ \midrule
\multirow{3}{*}{Toy, D=0.05}               & Enhanced EPR                              &   \textbf{0.054}$\pm${\scriptsize 0.020}        &  \textbf{0.048}$\pm${\scriptsize 0.019}         \\
                                           & HJB loss alone                                 &  {0.191}$\pm${\scriptsize 0.218}                                   & {0.160}$\pm${\scriptsize 0.179}                                 \\
                                           & Normalizing Flow                          &  {0.239}$\pm${\scriptsize 0.040}                                   & {0.174}$\pm${\scriptsize 0.034}                                   \\  \midrule
\multirow{3}{*}{Multi-stable}              & Enhanced EPR                              & \textbf{0.067}$\pm${\scriptsize 0.047}          & \textbf{0.066}$\pm${\scriptsize 0.049}     \\
                                           & HJB loss alone                                 & {0.249}$\pm${\scriptsize 0.015}                                  & {0.228}$\pm${\scriptsize 0.011}                                  \\
                                           & Normalizing Flow                          & {0.202}$\pm${\scriptsize 0.029}                                  & {0.133}$\pm${\scriptsize 0.017}                                  \\ \midrule
\multirow{3}{*}{Limit Cycle}               & Enhanced EPR                              & \textbf{0.070}$\pm${\scriptsize 0.016}        & \textbf{0.063}$\pm${\scriptsize 0.020}         \\
                                           & HJB loss alone                                 & {0.231}$\pm${\scriptsize 0.048}                                  &    {0.140}$\pm${\scriptsize 0.021}                                  \\
                                           & Normalizing Flow                          &  {0.384}$\pm${\scriptsize 0.021}                                  & {0.267}$\pm${\scriptsize 0.013}                                 \\  \bottomrule
\end{tabular}
\end{table*}

\begin{figure*}[!htbp]
\includegraphics[width=1.0\textwidth]{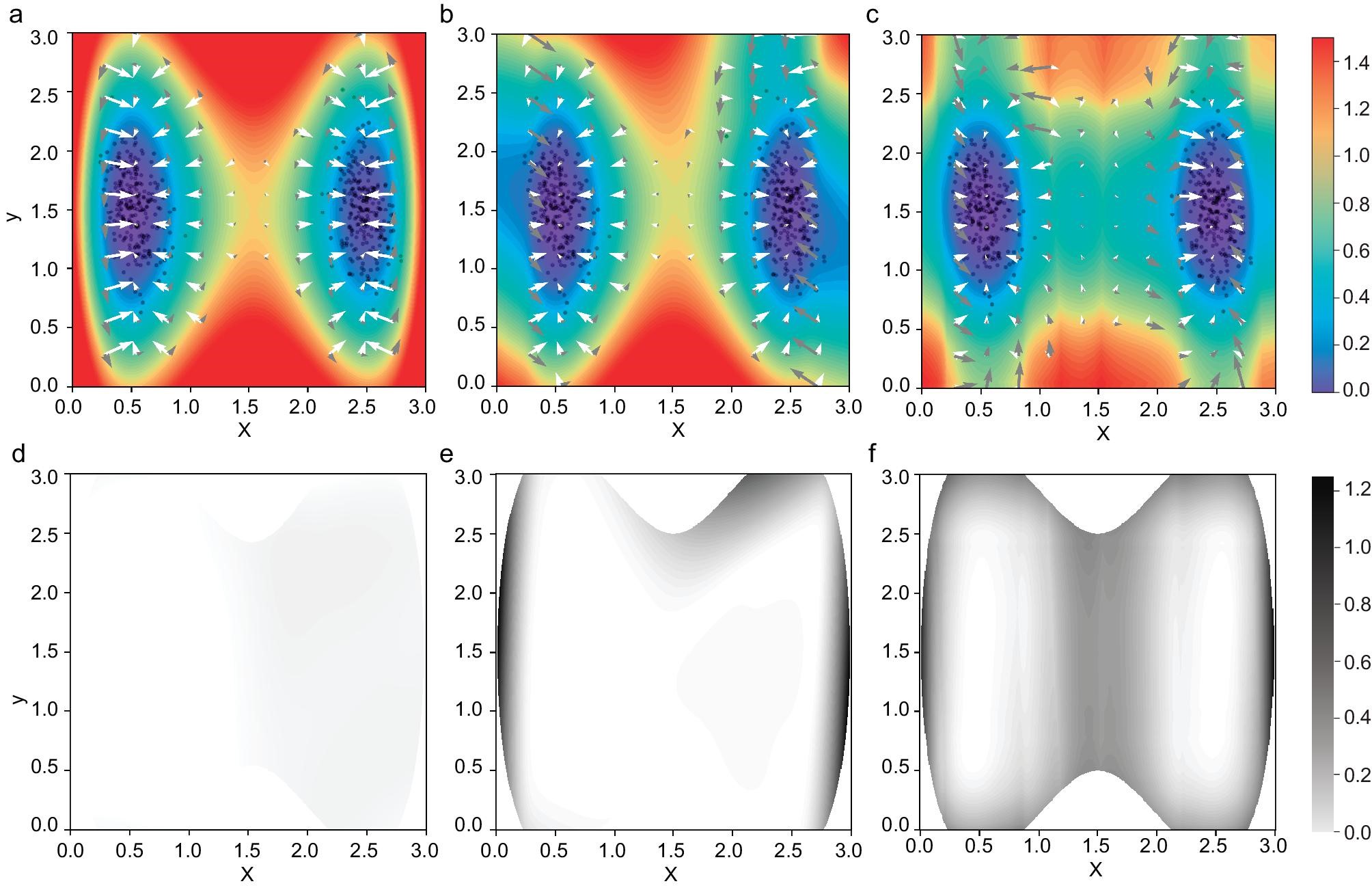}
    \caption{\textbf{Comparisons between models learned by (a,d) enhanced EPR, (b,e) HJB loss alone and (c,f) normalizing flow.} (a-c) Filled contour plots of the potential $V(\bs{x};\theta^*) $ for the toy model with $D=0.05$. The force field $\bs{F}(\bs{x})$ is decomposed into the gradient part $-\nabla V(\bs{x};\theta^*)$ (white arrows) and the non-gradient part (gray arrows). The length of an arrow denotes the scale of the vector. The solid dots are samples from the simulated invariant distribution. The results in the high-energy region $\{\bs{x} | V(\bs{x}) \geq 30D\}$ are omitted since they are less relevant to the dynamics. (d-f) The absolute error of the learned potential constructed in different ways.}
    \label{fig:2d_compare}
\end{figure*}

We present a detailed comparison of the 2D problems in Table~\textbf{\ref{tab:compare}}. Our experiments involve solving HJB alone with various distributions of enhanced samples $\mu(\bs x)$, and we report the optimal outcome as the representative entry for the table. For a fair comparison between enhanced EPR and standalone HJB, we maintain consistent network training configurations. Additionally, a detailed analysis of $\lambda_1$ and $\mu(\mathbf{x})$ in SI Section \ref{Sec F.1.} highlights the enhanced EPR's superior performance and robustness. Concerning the flow model, we leverage the same SDE-simulated dataset as utilized in enhanced EPR. The metrics presented in Table~{\ref{tab:compare}} yield the same consistent result, reinforcing the superiority of enhanced EPR over other methods across all problems.

We further illustrate the learned potential landscapes obtained through different methods for the 2D toy model with $D=0.05$ in Fig.~\ref{fig:2d_compare}a-c. Furthermore, we plot the absolute error between the learned potential and the reference solution in Fig.~\ref{fig:2d_compare}d-f. These plots demonstrate that errors within enhanced EPR are consistently small. In contrast, errors in HJB alone and NF can be significantly large, leading to unfavorable outcomes and limited smoothness. 
Based on this study and our numerical experiences, the advantages of the enhanced EPR over both the approach using HJB loss alone and NF can be summarized into the following points.

\begin{figure*}[h]
    \centering
    \includegraphics[width=0.75\textwidth]{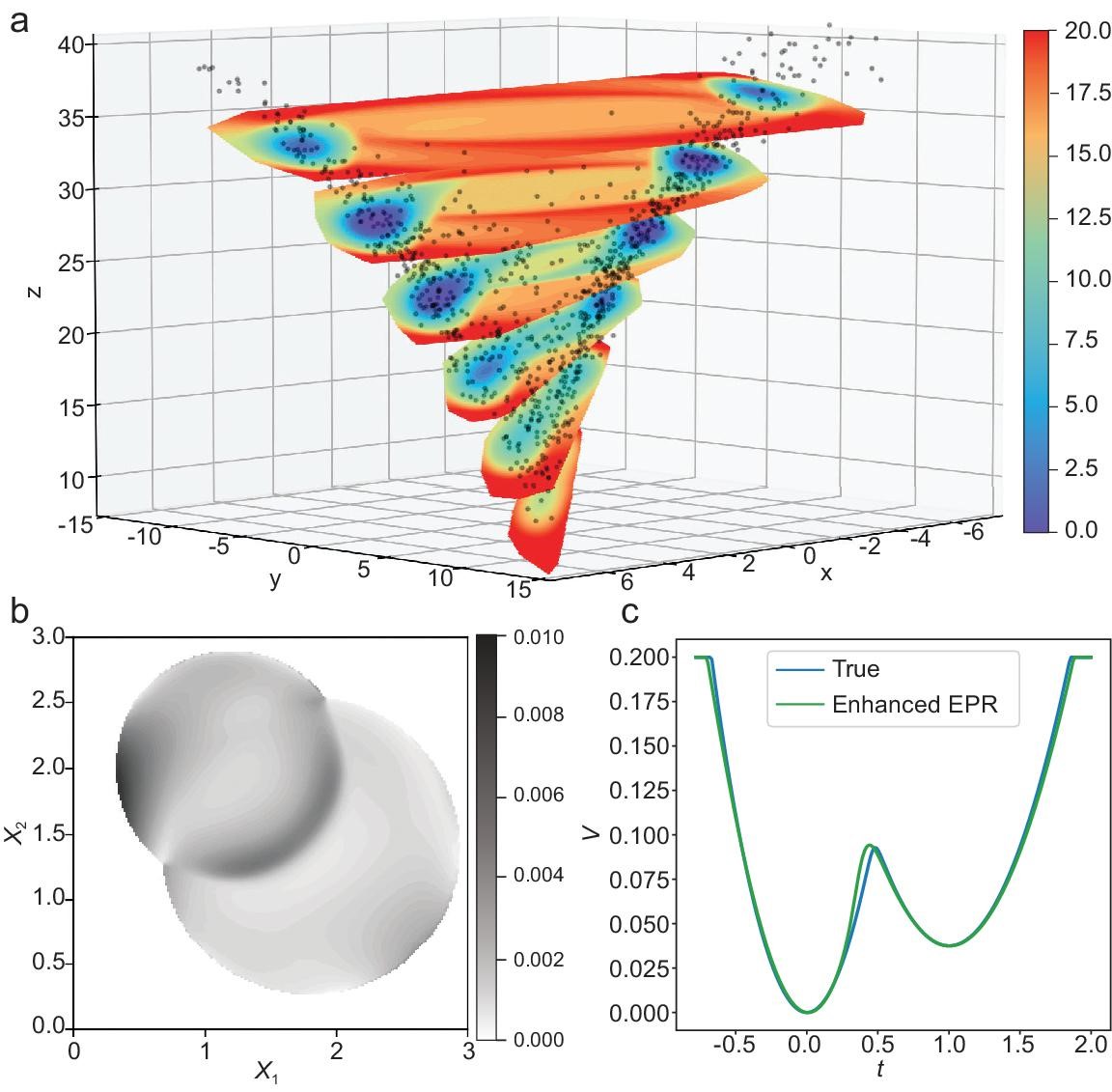}    
    \caption{\textbf{Landscapes constructed by enhanced EPR.} (a) Slices of the learned 3D potential ${V}(\bs{x};\theta^*)$ in the Lorenz system. The solid dots are samples from the simulated invariant distribution. (b) The accumulated measure of the absolute error between $U_0(\bs{x})$ and the learned solution $V(\bs{x}; \theta^*)$ based on the conditional probability $p_0(\bs{x} |x_1, x_2)$. (c) The learned 12D potential $V(\bs{x}; \theta^*)$ along a line $\bs{x}(t)=t \bs{\mu}_1 + (1-t) \bs{\mu}_2$, where $\bs{\mu}_1,\bs{\mu}_2$ are the means of two components in a 12D Gaussian mixture.}
    \label{fig:lorenz}
\end{figure*}

 \begin{itemize}
 \item \textit{Accuracy.} As shown in Table~\ref{tab:compare}, the enhanced EPR achieves the best accuracy over the HJB loss alone and normalizing flow approaches in terms of the rRMSE and rMAE metric. From Fig.~\ref{fig:2d_compare}, we can also find that the enhanced EPR presents the overall landscape more consistent with the simulated samples and the true/reference solution than the other two methods. The decomposition of the force also shows a better matching result. The learned potential by the HJB loss alone and normalizing flow tends to be rough and non-smooth near the edge of samples (Fig.~\ref{fig:2d_compare}b and c). Compared with the enhanced EPR and HJB loss alone, the normalizing flow captures the high probability domain but does not explicitly take advantage of the information of the dynamics, thus making its performance the worst.

 \item \textit{Robustness.} Without the guidance of EPR loss, minimizing HJB loss alone does not lead to a good approximation of the true solution that can closely match the heuristically chosen distribution as shown in Fig.~\ref{fig:2d_compare}b. The significant variance observed in the results for the toy model with $D=0.05$ by solving HJB alone is attributed to an exceptionally poor outlier. However, the enhanced EPR always gives reliable approximations with the introduced setup. This supports the robustness of the enhanced EPR loss. We refer to SI Section \ref{Sec F.1.} for detailed analysis of the chosen parameters $\lambda_1$ and distribution of enhanced samples $\mu(\bs x)$.
 
 \item \textit{Efficiency.} Our numerical experiments show that the enhanced EPR converges faster than the approach using the HJB loss alone. For instance, in the toy model with $D=0.05$, the enhanced EPR has achieved rRMSE of $0.088\pm0.083$ and rMAE of $0.075\pm0.070$ in 2000 epochs, while training with the HJB loss alone can not attain the same accuracy level even after 3000 epochs. As we have analyzed, this efficiency might be attributed to the strict convexity of the EPR loss. Related theoretical support for this speculation can be found in \cite{Mei2018}.
 
 \item \textit{Performance of single EPR.} We remark that
single EPR, which relies solely on the EPR loss by setting $\lambda=0$ in \eqref{eq:enh}, can still give acceptable results
with a relatively small $D = 0.05$. Its rRMSE and rMAE are $0.103\pm0.075$ and $0.087\pm0.063$, respectively, which are also small enough. In the limit cycle problem, single EPR is not as good as in the toy model and the multi-stable problem, since there are much fewer samples in the domain where the non-gradient force is extremely large, as shown in Fig.~\ref{fig:dw2d_force}b. However, single EPR loss can achieve competitive performance as long as the samples cover the domain effectively.
\end{itemize}

\subsection*{Three-dimensional Lorenz system}
We apply our landscape construction approach to the 3D Lorenz system~\cite{Lorenz63} with isotropic temporal Gaussian white noise, i.e. the system \eqref{eq:eq1} with 
$\bs{F}=(F_x,F_y,F_z)$ and
\begin{align}
F_x(x, y, z) & = \beta_1(y-x), \\
F_y(x, y, z) & =x\left(\beta_2-z\right)-y, \\
F_z(x, y, z) & =x y-\beta_3 z,
\end{align}
where $\beta_1=10, \beta_2=28$, and $\beta_3=\frac{8}{3}$. We add the noise with strength $D=1$.  This model was also considered in~\cite{Lin23} with $D=20$. We obtain the enhanced data $(\bs{x}'_i)_{1\le i\le N'}$ by adding Gaussian noises with standard deviation $\sigma=5$ to the SDE-simulated data $(\bs{x}_i)_{1\le i\le N}$, where $N=N'=10000$. We directly train the 3D potential $V(\bs{x};\theta)$ by enhanced EPR \eqref{eq:enh-loss-sm} with $\lambda_1=10.0, \lambda_2=1.0$, using Adam with a learning rate of 0.001 and batch size of 2048. A slice view of the landscape is presented in Fig.~\ref{fig:lorenz}a and the learned 3D potential agrees well with the simulated samples.

\subsection*{Twelve-dimensional multi-stable system}
To further validate the efficacy of EPR-Net in high-dimensional scenarios, we construct a Gaussian mixture model (GMM) in $\mathbb{R}^{12}$ with two centers, of which the true solution can be denoted as $U_0(\bs{x}) = -D \log p_0(\bs{x})$, where $p_0(\bs{x}) = w_1 p_1(\bs{x}) + w_2 p_2(\bs{x})$, and $p_i(\bs{x}) = Z_i^{-1}\exp(-(\bs{x}-\bs{\mu_i})^T \Sigma_i^{-1}(\bs{x}-\bs{\mu_i})/2), i=1, 2$, with normalization factor $Z_i$. The parameters are chosen as $w_1=0.6$, $w_2=0.4$, $\bs{\mu_1} = (1.2, 2.0, 0.6, 1.5, 0.9, 1.5, 1.5, 0.9, 1.2, 1.2, 0.5$, $ 1.8)^\top$, $\bs{\mu_2} = (1.8, 1.4, 0.8, 0.9, 0.9, 1.5, 2.0, 1.0$, $1.6, 1.0, 0.7, 1.4)^\top$, $\Sigma_1 = 0.04 I$, and $\Sigma_2 = 0.02 I$.

For evaluation, we first sample 10,000 points from the known GMM distribution and calculate the relative error between the learned 12D potential $V(\bs{x}; \theta^*)$ and the true potential $U_0(\bs{x})$. The resulting rRMSE and rMAE are 0.054 and 0.051, respectively, indicating the high accuracy of the learned potential. In addition, we denote the first two dimensions of this 12D problem as $(x_1, x_2)$ and draw samples from the corresponding conditional distribution $p_0(\bs{x} |x_1, x_2)$. As depicted in Fig.~\ref{fig:lorenz}b, we compute the cumulative measure of the absolute error between $U_0(\bs{x})$ and the learned solution $V(\bs{x})$ based on the conditional distribution as $\int |U_0(\bs{x}) - V(\bs{x})| p_0(\bs{x} |x_1, x_2) \dd \bs{x}$. Impressively, this integral consistently remains below 0.01, underscoring the fact that our approach yields negligible errors in the effective domain. Moreover, comparing the barrier heights (BHs) further validates the precision of our approach, with a relative error of less than 5\% for both BHs. Due to the diagonal covariance matrix, the saddle point precisely lies on the line passing through $\bs{\mu_1}$ and $\bs{\mu_2}$, allowing us to directly compare the BHs. As shown in Fig~\ref{fig:lorenz}c, the true and computed slice lines coincide well. For additional details, please refer to SI Section \ref{sec: F.5}.

\section*{DIMENSIONALITY REDUCTION}
\label{mt: dimen reduction}

When applying the approach above to high-dimensional problems, dimensionality reduction is necessary in order to visualize the results and gain physical insights. 
For simplicity, we consider the linear case and, with a slight abuse of notation, let $\bs{x} = (\bs{y}, \bs{z})^\top$, where $\bs{z} = (x_i, x_j) \in \mathbb{R}^2$ contains the coordinates of two variables of interest, and $\bs{y} \in \mathbb{R}^{d-2}$ corresponds to the remaining $d-2$ variables. The domain $\Omega$ (either $\mathbb{R}^d$ or a $d$-dimensional hyperrectangle) has the decomposition $\Omega=\Sigma \times \widetilde{\Omega}$, where  $\Sigma\subseteq \mathbb{R}^{d-2}$ and $\widetilde{\Omega}\subseteq \mathbb{R}^2$ are the domains of $\bs{y}$ and $\bs{z}$, respectively. 
Automatic selection of linear reduced variables and extensions to nonlinear reduced variables with general domains are also possible~\cite{zhang16, Kang21}.
In the current setting, the reduced potential is 
\be\label{eq:free-energy}
\widetilde{U}(\bs{z}) = -D\ln \widetilde{p}_{\ds}(\bs{z}) = -D\ln \int_{\Sigma} p_{\ds}(\bs{y}, \bs{z})\,\dd \bs{y},
\ee
and one can show that $\widetilde{U}$ minimizes the loss function:
\be\label{eq:P-EPR}
\operatorname{L_{P-EPR}}(\widetilde{V})=\int_{\Omega}\big|\bs{F}_{\bs{z}}(\bs{y}, \bs{z})+\nabla_{\bs{z}} \widetilde{V}(\bs{z};\theta)\big|^2 \,\dd \pi(\bs{y}, \bs{z}),
\ee
where $\bs{F}_{\bs{z}}(\bs{y}, \bs{z})\in \mathbb{R}^{2}$ is the $\bs{z}$-component of the force field $\bs{F}=(\bs{F}_{\bs{y}}, \bs{F}_{\bs{z}})^\top$.

Moreover, one can derive an enhanced loss as in \eqref{eq:enh} that could be used for systems with small $D$.  
To this end, we note that $\widetilde{U}$ satisfies the projected HJB equation
\begin{align}\label{eq:P-HJB}
\mathcal{N}_{\textrm{P-HJB}}(\widetilde{U}):= &  -\widetilde{\bs{F}}  \cdot \nabla_{\bs{z}} \widetilde{U}  + D \Delta_{\bs{z}} \widetilde{U} \nonumber \\
& -|\nabla_{\bs{z}} \widetilde{U}|^2
 + D \nabla_{\bs{z}} \cdot \widetilde{\bs{F}} =0\,,
\end{align}
with asymptotic BC $\widetilde{U}\rightarrow \infty$ as $|\bs{z}|\rightarrow \infty$, or the reflecting BC $(\widetilde{\bs{F}}+\nabla_{\bs{z}} \widetilde{U}) \cdot \widetilde{\bs{n}}=0$ on $\partial \widetilde{\Omega}$,
where $\widetilde{\bs{F}}(\bs{z}):=\int_{\Sigma} \bs{F}_{\bs{z}}(\bs{y}, \bs{z}) \dd \pi(\bs{y}|\bs{z})$ is the projected force defined using the conditional distribution $\dd \pi(\bs{y}|\bs{z}) = p_{\ds}(\bs{y},\bs{z})/\widetilde{p}_{\ds}(\bs{z})\,\dd\bs{y}$, and $\widetilde{\bs{n}}$ denotes the unit outer normal on $\partial\widetilde{\Omega}$. 
Based on \eqref{eq:P-HJB}, we can formulate the projected HJB loss 
\begin{equation}\label{eq:l-phjb}
\operatorname{L_{P-HJB}}(\widetilde{V})=\int_{\widetilde{\Omega}} \big|\mathcal{N}_{\textrm{P-HJB}}(\widetilde{V})\big|^2 \,\dd \mu(\bs{z}),
\end{equation}
where $\mu$ is any suitable distribution over $\widetilde{\Omega}$, and $\widetilde{\bs{F}}$ in \eqref{eq:P-HJB} is learned beforehand by training a DNN with the loss
\begin{equation}\label{eq:P-For-Loss}
\operatorname{L_{P-For}}(\widetilde{\bs{G}})=\int_{\Omega}\big|\bs{F}_{\bs{z}}(\bs{y}, \bs{z})- \widetilde{\bs{G}}(\bs{z}; \theta)\big|^2 \, \dd \pi(\bs{y}, \bs{z}).
\end{equation}
 The overall enhanced loss \eqref{eq:enh-loss-sm} used in numerical computations comprises two terms, which are empirical estimates of \eqref{eq:P-EPR} and \eqref{eq:l-phjb} based on two different sets of sample data. We refer the reader to SI Section \ref{Sec D: Dimensionality reduction} for derivation details, SI Section~\ref{sec: G.1} for experiments of Ferrell's cell cycle model~\cite{Ferrell11} and SI Section \ref{sec: G.2} and \ref{sec: G.3} for details and additional results of the 8D~\cite{Wang10} and 52D~\cite{Li13} models.

\begin{figure*}[!htbp]
    \centering
    \includegraphics[width=1.0\textwidth]{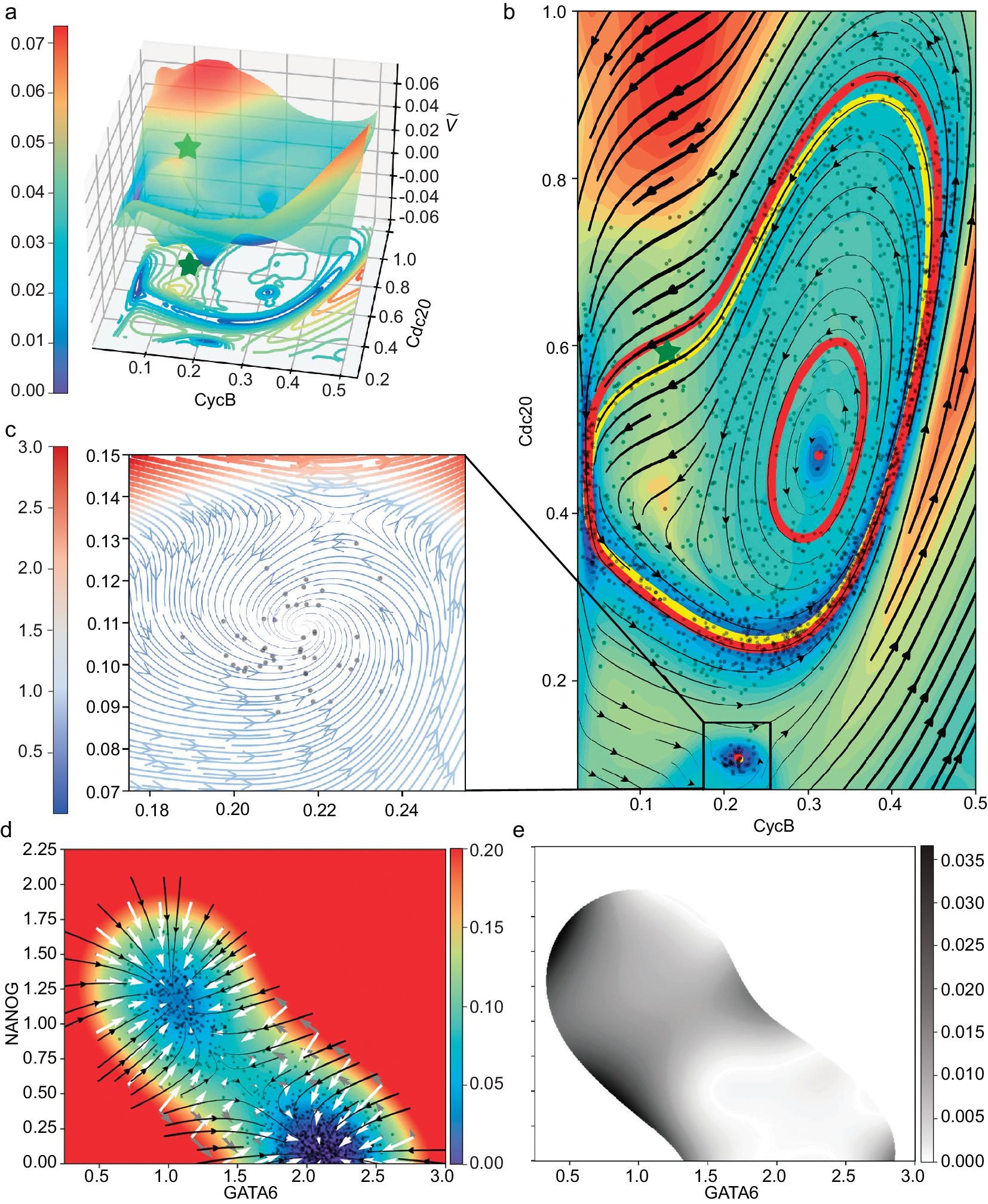}
    \caption{\textbf{Dimensionality reduction of high-dimensional systems.} (a-c) Eight-dimensional cell cycle model. (a) Reduced potential landscape $\widetilde{V}$ with projected contour lines. The green star at $(0.13,0.59)$  denotes the saddle point of $\widetilde{V}$. (b) Projected sample points, streamlines of the projected force field $\widetilde{\bs{G}}(\bs{x};\theta^*)$ and the filled contour plot of $\widetilde{V}(\bs{x};\theta^*)$. Two red circles and two red dots (close to $(0.22,0.11)$ and $(0.31,0.47)$, respectively) show the stable limit sets of the projected force field. The yellow circle is the projection of the original high-dimensional limit cycle. (c) An enlarged view of the square domain in (b), showing the detailed spiral structure of the streamlines of $\widetilde{\bs{G}}(\bs{x};\theta^*)$ around the stable point. (d and e) Fifty-two-dimensional multi-stable system. (d) Projected force $\widetilde{\bs{G}}(\bs{z};\theta^*)$ and potential $\widetilde{V_1}(\bs{z};\theta^*)$ of the 52D double-well model learned by enhanced EPR. (e) The absolute error of the reduced potential constructed in different ways, \textit{i.e.}, $|\widetilde{V_1}(\bs{z};\theta^*) - \widetilde{V_2}(\bs{z};\theta^*)|$.}
    \label{fig:dr}
\end{figure*}

\subsection*{Eight-dimensional cell cycle: reduced potential} \label{sec: exp 8d}
Subsequently, we apply our dimensionality reduction approach to construct the landscape for an 8D cell cycle model \cite{Wang10}, which contains both a limit cycle and a stable equilibrium point. In this case, we consider CycB (a cyclin B protein) and Cdc20 (an exit protein) as the reduced variables following \cite{Wang10}. As shown in Fig.~\ref{fig:dr}a and b, we can find that both the profile of the reduced potential and the force strength agree well with the density of projected samples. Moreover, we gain some important insights from Fig.~\ref{fig:dr}b on the projection of the high-dimensional dynamics to two dimensions. One particular feature is that the limit cycle induced by the projected force $\widetilde{\bs{G}}(\bs{x};\theta^*)$ (outer red circle) slightly differs from the limit cycle directly projected from high dimensions (yellow circle), and the difference is either minor or moderate depending on whether the sample density near the limit circle is high or low. This is natural in the reduction when $D>0$, since the distribution $\pi(\bs{y}|\bs{z})$ involved in computing $\widetilde{\bs{G}}(\bs{x};\theta^*)$ is not a Dirac $\delta$-distribution, but a diffusive distribution with varying widths along the limit cycle, and the difference will disappear as $D\rightarrow 0$. Another feature is that we unexpectedly get an additional stable limit cycle (inner red circle) and a stable point (red dot in the center) emerging inside the outer limit cycle. Though virtual in high dimensions and biologically irrelevant, the existence of such two limit sets is reminiscent of the Poincar\'e-Bendixson theorem in planar dynamics theory~\cite[Chapter 10.6]{Hirsch04}, which depicts a common phenomenon when performing dimensionality reduction with limit cycles to the 2D plane. Note that this theorem does not guarantee the stability of such fixed points; they could be stable, unstable, or even saddle points, see another case in SI \ref{sec: G.1}. The emergence of these two limit sets is specific in this model due to the relatively flat landscape of the potential in the centering region.  In addition, close to the saddle point of $\widetilde{V}$ (green star), there is a barrier along the limit cycle direction, while there is a local well along the Cdc20 direction, which characterizes the region that biological cycle paths mainly go through.  Last but not least, an enlarged view of the local attractive domain outside the limit cycle shows its intricate spiral structure  (Fig.~\ref{fig:dr}c), which has not been revealed in previous work based on mean-field approximation~\cite{Wang10}.

\subsection*{Fifty-two-dimensional multi-stable system: high-dimensional and reduced potentials} We compare two approaches to construct the reduced potential. One is to learn the reduced force $\widetilde{\bs{G}}(\bs{z}; \theta^*)$ by~\eqref{eq:P-For-Loss} first and then use it to construct the landscape $\widetilde{V}(\bs{z}; \theta^*)$ by enhanced EPR in $\bs{z}$-space. The other is to build a high-dimensional potential $V(\bs{x}; \theta^*)$ by enhanced EPR first and then reduce it (see SI Section \ref{Sec D.1: Dimensionality reduction}). For this comparison, we apply our approach to a gene stem cell regulatory network described by an ODE in 52 dimensions~\cite{Li13} and take GATA6 (a major differentiation marker gene) and NANOG (a major stem cell marker gene) as the reduced variable $\bs{z}$, as suggested in~\cite{Li13}. We define $\mathcal{A}_i$ as the set of indices for activating $x_i$ and $\mathcal{R}_i$ as the set of indices for repressing $x_i$, the corresponding relationships are defined as the 52D node network shown in~\cite{Li13}. For $i = 1, ..., 52$, 
\begin{equation}
F_i(\bs{x}) = -k x_i+\sum_{j\in\mathcal{A}_j}^{} \frac{a x_j^n}{S^n+x_j^n}+\sum_{j\in\mathcal{R}_j}^{} \frac{b S^n}{S^n+x_j^n},
\end{equation}
where $a=0.37$, $b=0.5$, $k=1$, $S=0.5$, and $n=3$. We choose the noise strength $D=0.02$ and focus on the domain $\Omega=[0, 3]^{52}$. 

As shown in Fig.~\ref{fig:dr}d, the projected force $\widetilde{\bs{G}}(\bs{z};\theta^*)$ demonstrates the reduced dynamics and the constructed potential $\widetilde{V_1}(\bs{z};\theta^*)$ agrees well with the SDE-simulated samples. While $\widetilde{V_1}(\bs{z};\theta^*)$ is learned by first obtaining the projected force, $\widetilde{V_2}(\bs{z};\theta^*)$ is reduced from a high-dimensional ${V}(\bs{x};\theta^*)$ precomputed by enhanced EPR.
The gray plot of the absolute error of $|\widetilde{V_1}(\bs{z};\theta^*) - \widetilde{V_2}(\bs{z};\theta^*)|$ is shown in Fig.~\ref{fig:dr}e, which supports the consistency of two potentials learned by different approaches. When taking $\widetilde{V_1}(\bs{z};\theta^*)$ as the reference solution, we get the rRMSE 0.113 and rMAE 0.097. The minor relative errors show that these two approaches to constructing the reduced potential are quantitatively consistent, although it is difficult to know which is more accurate. It also indirectly supports the reliability of the learned high-dimensional potential $V(\bs{x}; \theta^*)$. The obtained landscape shows a smoother and more delicate profile compared to the mean-field approach \cite{Li13}.

\section*{DISCUSSION AND CONCLUSION}

The EPR-Net formulation can be extended to the case of state-dependent diffusion coefficients without difficulty. Consider the It\^o SDE
\be\label{eq:SDE-Var}
\frac{\dd \bs{x}(t)}{\dd t}=\bs{F}(\bs{x}) + \sqrt{2D}\sigma(\bs{x})\,\dot{\bs{w}},\ \bs{x}(0)=\bs{x}_0,
\ee
with diffusion matrix $\sigma(\bs{x})\in\mathbb{R}^{d\times m}$ and an $m$-dimensional temporal Gaussian white noise $\dot{\bs{w}}$. We assume that $m\ge d$ and the matrix $a(\bs{x}):=(\sigma\sigma^\top)(\bs{x})$ satisfies  
$\bs{u}^\top a(\bs{x})\bs{u} \ge c_0 |\bs{u}|^2$ for all $\bm{x}, \bs{u}\in \mathbb{R}^d$, where $c_0>0$ is a positive constant. Using a similar derivation as before, 
we can again show that the high-dimensional landscape function $U$ of \eqref{eq:SDE-Var} minimizes the EPR loss 
\begin{equation}\label{eq:EPR-Var}
\operatorname{L_{V-EPR}}(V) = \int_{\Omega}\left|\bs{F}^v(\bs{x})+ a(\bs{x}) \nabla V(\bs{x})\right|_{a^{-1}}^2 \,\dd \pi{(\bs{x})},
\end{equation}
where $\bs{F}^v(\bs{x}) = \bs{F}(\bs{x}) - D \nabla \cdot a(\bs{x})$ and $|\bs{u}|_{a^{-1}}^2:=\bs{u}^\top a^{-1}(\bs{x})\bs{u}$ for $\bs{u}\in \mathbb{R}^d$. 
We provide derivation details of \eqref{eq:EPR-Var} in SI Section \ref{Sec E: coefficients}, and we leave the numerical study of \eqref{eq:SDE-Var}--\eqref{eq:EPR-Var} for future work.  

% \subsection{Conclusions}

To summarize, we have demonstrated the formulation, applicability and superiority of EPR to the benchmark and real biological examples over some existing approaches. Below we make some final remarks. First, concerning the use of the steady-state distribution $\pi(\bs{x})$ in \eqref{eq:L-EPR} and its approximation by a long time series of the SDE \eqref{eq:eq1} in EPR-Net, we emphasize that it is the sampling approximation of $\pi$ that naturally captures the important parts of the potential function, and the landscape beyond the sampled regions is not that essential in practice. Second, as is exemplified in Fig.~\ref{fig:dw2d_force} and Table~\ref{tab:compare}, we found that a direct application of density estimation methods (DEM), e.g., normalizing flows \cite{Koby21}, to the sampled time series data does not give potential landscape with satisfactory accuracy. We speculate that such deficiency of DEM is due to its over-generality and the fact that it does not take advantage of the force field information explicitly compared to \eqref{eq:L-EPR}. Finally, we remark that in our dimensionality reduction approach, we choose the reduced variables according to available biophysics prior knowledge. Our approach also exhibits the capability to be extended to incorporate a state-dependent diffusion coefficient. Automatic selection of general nonlinear reduced variables can be also studied by incorporating the idea of the identification of collected variables in molecular simulations \cite{Parr2019,Parr2020}. 

Overall, we have presented the EPR-Net, a simple yet effective DNN approach, for constructing the non-equilibrium potential landscape of NESS systems. This approach is both elegant and robust due to its variational structure and 
 its flexibility to be combined with other types of loss functions. Further extension of dimensionality reduction to nonlinear reduced variables and numerical investigations in the case of state-dependent diffusion coefficients will be explored in future works. Moreover, we are actively considering the application of our method to single-cell RNA sequencing data, which involves systems of really high dimensionality and typically lacks analytical ODE formulations. We will explore this point in our forthcoming studies.

\section*{METHODS}
Detailed methods are given in the supplementary material. The code is accessible at \href{https://github.com/yzhao98/EPR-Net}{https://github.com/yzhao98/EPR-Net}.

\section*{ACKNOWLEDGMENTS}
We thank Professors Chunhe Li, Xiaoliang Wan and Dr.\ Yufei Ma for helpful discussions.  The computation of this work was conducted on the High-performance Computing Platform of Peking University. 

\section*{FUNDING}
T.L. and Y.Z. acknowledge support from the National Natural Science Foundational of China (11825102 and 12288101) and the Ministry of Science and Technology of China (2021YFA1003301). The work of W.Z. was supported by the Deutsche Forschungsgemeinschaft (DFG) under Germany’s Excellence Strategy, part of the Berlin Mathematics Research Centre MATH+ (EXC-2046/1, 390685689), and the DFG through Grant CRC 1114 `Scaling Cascades in Complex Systems' (235221301).

\section*{AUTHOR CONTRIBUTIONS}
Y.Z., W.Z. and T.L. designed the research; Y.Z. performed the research; Y.Z., W.Z. and T.L. wrote the paper.

\section*{}

\noindent \textbf{\textit{Conflict of interest statement.}} None declared.

\appendix

\bibliographystyle{nsr}
\bibliography{ref}

\onecolumn

\vspace*{2em} 
\begin{center} 
\Large 
\textbf{Supplementary Information for EPR-Net:\\ Constructing non-equilibrium potential landscape via a variational force projection formulation}
\end{center}
\vspace*{1em}

\tableofcontents

\setcounter{section}{0} 
\setcounter{secnumdepth}{3}
\renewcommand{\thesection}{\Alph{section}}
\renewcommand{\thesubsection}{\Alph{section}.\arabic{subsection}}
  
\newpage 

In this supplementary information (SI), we will present further theoretical derivations and computational details of the contents in the main text (MT). 
This SI consists of two parts: \textit{theory} and \textit{computation}.

\vspace{4mm}
\addcontentsline{toc}{section}{\bf Part 1: Theory}
\vspace*{2mm}
\noindent{\color{blue} \textbf{PART 1: THEORY}}

\vspace*{4mm}

We will first provide details of theoretical derivations omitted in the MT.

\section{Derivation of the EPR loss}
\label{appsec A:validation-EPR}
In this section, we show that, up to an additive constant, the potential function $U(\bs{x}):= -D\ln p_{\ds}(\bs{x})$ is the unique minimizer of the EPR loss in equation (1) in MT.

First, we show that the orthogonality relation 
\be\label{appeq:Orth}
\int_{\Omega}(\bs{F}+\nabla U)\cdot \nabla W\, \dd \pi = 0
\ee
holds for any suitable function $W(\bs{x}):\mathbb{R}^d\rightarrow \mathbb{R}$ under both choices of the boundary conditions considered in this work, where $\dd \pi(\bs{x}):= p_{\ds}(\bs{x}) \dd \bs{x}$. To see this, we note that
\begin{align*}
 & \int_{\Omega}(\bs{F}+\nabla U)\cdot \nabla W\, \dd \pi\\
 = &\int_{\Omega} (p_{\ds}\bs{F}-D\nabla p_{\ds})\cdot\nabla W\,  \dd \bs{x}\\
 = &\int_{\partial\Omega} W (p_{\ds}\bs{F}-D\nabla p_{\ds})\cdot \bs{n}  \, \dd \bs{x}\\
  & -\int_{\Omega} W \nabla\cdot (p_{\ds}\bs{F}-D\nabla p_{\ds})  \, \dd \bs{x}\\
  = & 0
\end{align*}
where we have used integration by parts, the relation $p_{\ds}(\bs{x})= \exp(-U(\bs{x})/D)$, the corresponding BC, and the steady state Fokker-Planck equation (FPE) satisfied by $p_{\ds}$.

Now consider the EPR loss, we have
\begin{align*}
\operatorname{L_{EPR}}(V)= & \int_{\Omega}|\bs{F}+\nabla V|^2 \,\dd \pi\\
=& \int_{\Omega}|\bs{F}+\nabla U|^2 +  |\nabla V-\nabla U|^2 \,\dd \pi,
\end{align*}
where we have used the orthogonality relation \eqref{appeq:Orth} to arrive at the last equality, from which we conclude that $U(\bs{x})$ is the unique minimizer of the EPR loss up to an additive constant.

\section{Stability of the EPR minimizer}\label{Sec B: stability}

In this section, we formally show that small perturbations of the invariant distribution $\pi$ will not introduce a disastrous change to the minimizer of the corresponding EPR loss. We only consider the case where $\Omega$ is a $d$-dimensional hyperrectangle. The argument for $\Omega=\mathbb{R}^d$ is similar. 

Suppose $\dd\pi(\xb)=p(\xb)\dd\xb$, $\dd\mu(\xb)=q(\xb)\dd\xb$, and the functions $U(\xb)$ and $\bar{U}(\xb)$ are the unique minimizers (up to a constant) of the following two EPR losses
\begin{align*}
U=\mathop{\arg\min}_{V} \int_\Omega \left|\bs{F} + \nabla V \right|^2\,  \dd \pi,\\
\bar{U}=\mathop{\arg\min}_{V} \int_\Omega \left|\bs{F} + \nabla V \right|^2\, \dd \mu,
\end{align*}
respectively. It is not difficult to find that the Euler-Lagrange equations of $U,\bar{U}$ are given by the following partial differential equation (PDE) with suitable BCs:
\begin{align*}
\nabla\cdot((\bs{F}+\nabla U)p)=0\mbox{ in }\Omega,\ (\bs{F}+\nabla U)\cdot\bs{n}=0\mbox{ on }\partial\Omega,\\    
\nabla\cdot((\bs{F}+\nabla \bar{U})q)=0\mbox{ in }\Omega,\ (\bs{F}+\nabla \bar{U})\cdot\bs{n}=0\mbox{ on }\partial\Omega.    
\end{align*}
The PDEs above defined inside the domain $\Omega$ can be converted to 
\begin{align*}
\Delta U p + \nabla U\cdot\nabla p = -\nabla\cdot(p\bs{F}),\\
\Delta \bar{U} q + \nabla \bar{U}\cdot\nabla q = -\nabla\cdot(q\bs{F}).
\end{align*}
Define $U_0(\xb)=-D\ln p(\xb)$ and $\bar{U}_0(\xb)=-D\ln q(\xb)$. We then obtain
\begin{align}
- \nabla U\cdot\nabla U_0 + D\Delta U  = \bs{F}\cdot \nabla U_0-D\nabla\cdot \bs{F},\label{appeq:U1}\\
- \nabla \bar{U}\cdot\nabla \bar{U}_0 + D\Delta \bar{U}  = \bs{F}\cdot \nabla \bar{U}_0-D\nabla\cdot \bs{F}.\label{appeq:U2}
\end{align}
Assuming that $\delta U_0:= U_0-\bar{U}_0=O(\varepsilon)$, where $0 < \epsilon \ll 1$ denotes a small constant, we have the PDE for $U-\bar{U}$ by subtracting \eqref{appeq:U2} from \eqref{appeq:U1}:
\begin{align*}\label{appeq:U-Diff}
-\nabla (U- & \bar{U})\cdot \nabla U_0 + D\Delta(U-\bar{U}) = \bs{F}\cdot \nabla (\delta U_0)+\nabla \bar{U}\cdot \nabla (\delta U_0)
\end{align*}
with BC $\nabla (U-\bar{U})\cdot \bs{n}=0$. Since $U_0,\bar{U},\bs{F}\sim O(1)$, we can obtain that 
\begin{equation*}
U(\xb)-\bar{U}(\xb)=O(\varepsilon)
\end{equation*}
by the regularity theory of elliptic PDE \cite[Section~6.3]{Evans10} when $D\sim O(1)$, or by the matched asymptotic expansion  when $D\ll 1$ \cite[Chapter~2]{Holmes13}. In fact, the closeness between $U(\bs{x})$ and $\bar{U}(\bs{x})$ can be ensured as long as $U_0$ and $\bar{U}_0$ are close enough in the region where $p(\bs{x})$ and $q(\bs{x})$ are bounded away from zero by the method of characteristics analysis \cite[Section~2.1]{Evans10} and matched asymptotics.

The above derivations assume that the PDFs $p$,$q$ are smooth enough. The stability analysis for general distributions $\pi$ and $\mu$ is also possible by utilizing the functional analysis tools. However, the derivations  will be abstract and quite involved, and we will report it elsewhere.

\section{Dimensionality reduction}\label{Sec D: Dimensionality reduction}

%%%%%%%%%%%%%%%%%%%%%%%%%%

In this section, we study dimensionality reduction for high-dimensional problems in order to learn the projected potential. A straightforward approach is to first learn the high-dimensional potential $U$ and then find its low-dimensional representation, i.e., the reduced potential or the free energy function, using dimensionality reduction techniques. An alternative approach is to directly learn the low-dimensional reduced potential.

Denote by $\bs{x} = (\bs{y}, \bs{z})^\top\in \Omega$. As in MT Section 1.4, we assume the domain 
$$\Omega=\Sigma \times \widetilde{\Omega},$$ 
where $\Sigma\subseteq \mathbb{R}^{d-2}$ and $\widetilde{\Omega}\subseteq \mathbb{R}^2$ are the domain of $\bs{y}$ and $\bs{z}$, respectively. The reduced potential $\widetilde{U}(\bs{z})$ is defined as 
\be\label{appeq:free-energy}
\widetilde{U}(\bs{z}) = -D\ln \widetilde{p}_{\ds}(\bs{z}) = -D\ln \int_{\Sigma} p_{\ds}(\bs{y}, \bs{z})\,\dd \bs{y}.
\ee

One natural approach for constructing $\widetilde{U}(\bs{z})$ is directly integrating 
$p_{\ds}(\bs{y}, \bs{z})$ based on the learned $U(\bs{y}, \bs{z})$ with the EPR loss, i.e.,
\be
\widetilde{U}(\bs{z}) = -D\ln \int_{\Sigma} \exp(-U(\bs{y}, \bs{z})/D)\,\dd \bs{y}.
\ee
However, performing this integration is not a straightforward numerical task (see, e.g., \cite[Chapter 7]{Frenkel02}). 

\subsection{Gradient projection loss} \label{Sec D.1: Dimensionality reduction}

In this subsection, we study a simple approach to approximate $\widetilde{U}(\bs{z})$ based on sample points, which approximately obey the invariant distribution $\pi(\xb)$, and the learned high dimensional potential function $U(\bs{x})$ by EPR loss. This approach is taken in the consistency checking of the reduced potentials by different methods in SI Section~\ref{sec: G.3}.  The idea is to utilize the gradient projection (GP) loss on the $\bs{z}$ components of $\nabla U$:
\be\label{appeq:GP-Loss}
\operatorname{L_{GP}}(\widetilde{V})=\int_{\Omega}\big|\nabla_{\bs{z}} U(\bs{y},\bs{z})-\nabla_{\bs{z}}\widetilde{V}(\bs{z})\big|^2\, \dd \pi(\bs{y}, \bs{z}).
\ee
To justify \eqref{appeq:GP-Loss}, we note that
\begin{align*}
\operatorname{L_{GP}}(\widetilde{V})  = & \int_{\Omega}\big|\nabla_{\bs{z}} U-\nabla_{\bs{z}}\widetilde{V}\big|^2 \,\dd \pi(\bs{x})\\
= & \int_{\Omega}\big|\nabla_{\bs{z}} U-\nabla_{\bs{z}}\widetilde{U}+\nabla_{\bs{z}}\widetilde{U}-\nabla_{\bs{z}}\widetilde{V}\big|^2\, \dd \pi(\bs{x})\\
=& \int_{\Omega}\Big(\big|\nabla_{\bs{z}} U-\nabla_{\bs{z}}\widetilde{U}\big|^2 + \big|\nabla_{\bs{z}}\widetilde{U}-\nabla_{\bs{z}}\widetilde{V}\big|^2\Big)\, \dd \pi(\bs{x}) \\
&+2\int_{\Omega} (\nabla_{\bs{z}} U-\nabla_{\bs{z}}\widetilde{U})\cdot \nabla_{\bs{z}}(\widetilde{U}-\widetilde{V})\,\dd \pi(\bs{x})\\
=: & P_1+P_2,
\end{align*}
where $P_1$ and $P_2$ denote the terms in the third and the fourth line above, respectively. 
The term $P_2=0$ since
\begin{align*}
 &\int_{\Omega} \nabla_{\bs{z}} U\cdot \nabla_{\bs{z}}(\widetilde{U}-\widetilde{V})\,\dd \pi(\bs{x})\\
=& \int_{\widetilde{\Omega}} \left( \int_{\Sigma}\nabla_{\bs{z}} U  \mathrm{e}^{-\frac{U}{D}} \dd \bs{y}\right)  \cdot \nabla_{\bs{z}}(\widetilde{U}-\widetilde{V}) \,\dd \bs{z}\\
=& -D\int_{\widetilde{\Omega}} \nabla_{\bs{z}}\left( \int_{\Sigma} \mathrm{e}^{-\frac{U}{D}} \,\dd \bs{y}\right)  \cdot \nabla_{\bs{z}}(\widetilde{U}-\widetilde{V}) \,\dd \bs{z}\\
=& -D\int_{\widetilde{\Omega}} \nabla_{\bs{z}}\widetilde{p}_{\ds}  \cdot \nabla_{\bs{z}}(\widetilde{U}-\widetilde{V}) \,\dd \bs{z}\\
=& \int_{\widetilde{\Omega}} \nabla_{\bs{z}}\widetilde{U}  \cdot \nabla_{\bs{z}}(\widetilde{U}-\widetilde{V}) \,\widetilde{p}_{\ds} \,\dd \bs{z}
\end{align*}
and 
\begin{align*}
 &\int_{\Omega} \nabla_{\bs{z}}\widetilde{U} \cdot \nabla_{\bs{z}}(\widetilde{U}-\widetilde{V})\,\dd \pi(\bs{x})
=  \int_{\widetilde{\Omega}} \nabla_{\bs{z}}\widetilde{U} \cdot \nabla_{\bs{z}}(\widetilde{U}-\widetilde{V})\,\widetilde{p}_{\ds} \,\dd \bs{z}, 
\end{align*}
which cancel with each other in $P_2$. 

Therefore, the minimization of GP loss is equivalent to minimizing
$$\int_{\widetilde{\Omega}}\big|\nabla_{\bs{z}}\widetilde{U}-\nabla_{\bs{z}}\widetilde{V}\big|^2\, \widetilde{p}_{\ds}\, \dd \bs{z},$$ which clearly implies that $\widetilde{U}(\bs{z})$ is the unique minimizer (up to a constant) of the proposed GP loss.

\subsection{Projected EPR loss} \label{Sec D.2: Dimensionality reduction}
\label{subsec:P-EPR}
In this subsection, we study the projected EPR (P-EPR) loss, which has the form
\be\label{appeq:P-EPR}
\operatorname{L_{P-EPR}}(\widetilde{V})=\int_{\Omega}\big|\bs{F}_{\bs{z}}(\bs{y}, \bs{z})+\nabla_{\bs{z}} \widetilde{V}(\bs{z})\big|^2\, \dd \pi(\bs{y}, \bs{z}),
\ee
where $\bs{F}_{\bs{z}}(\bs{y}, \bs{z})\in \mathbb{R}^{2}$ is the $\bs{z}$-component of the force field $\bs{F}=(\bs{F}_{\bs{y}}, \bs{F}_{\bs{z}})^\top$.

Define
\be\label{appeq:P-EPR-T}
\operatorname{\widetilde{L}_{P-EPR}}(\widetilde{V})= \int_{\Omega}\big|\bs{F}(\bs{y},\bs{z})+\nabla \widetilde{V}(\bs{z})\big|^2\, \dd \pi(\bs{y},\bs{z}),
\ee
where $\nabla$ is the full gradient with respect to $\xb$. To justify \eqref{appeq:P-EPR}, 
we first note the following equivalence
\be\label{appeq:P-EPR-Equi1}
\min \operatorname{L_{P-EPR}}(\widetilde{V})\quad \Longleftrightarrow\quad 
\min \operatorname{\widetilde{L}_{P-EPR}}(\widetilde{V}),
\ee
since $\nabla_{\bs{y}}\widetilde{V}(\bs{z})=0$ and the $\bs{y}$-components of $\bs{F}+\nabla \widetilde{V}$ only introduce an irrelevant constant in \eqref{appeq:P-EPR-T}. Furthermore, we have 
\begin{align*}
\operatorname{\widetilde{L}_{P-EPR}}(\widetilde{V})= & \int_{\Omega}\big|\bs{F}+\nabla \widetilde{V}\big|^2 \,\dd \pi(\bs{x})\\
= & \int_{\Omega}\big|\bs{F}+\nabla U +\nabla \widetilde{V} - \nabla U\big|^2 \,\dd \pi(\bs{x})\\
= & \int_{\Omega}\big|\bs{F}+\nabla U\big|^2 + \big|\nabla \widetilde{V} - \nabla U\big|^2 \,\dd \pi(\bs{x}),
\end{align*}
where the last equality is due to the orthogonality relation \eqref{appeq:Orth}. 

Using a similar argument for deriving \eqref{appeq:P-EPR-Equi1}, the equivalence \eqref{appeq:P-EPR-Equi1} itself, as well as the GP loss in \eqref{appeq:GP-Loss}, we get
\be
\min \operatorname{L_{P-EPR}}(\widetilde{V})\quad \Longleftrightarrow\quad 
\min \operatorname{L_{GP}}(\widetilde{V}).
\ee
Since $\widetilde{U}$ minimizes the GP loss as is shown in the previous subsection, we conclude that $\widetilde{U}$ minimizes the loss in \eqref{appeq:P-EPR}. 

\subsection{Force projection loss}  \label{Sec D.3: Dimensionality reduction}

In this subsection, we study the force projection (P-For) loss for approximating the projection of $\bs{F}_{\bs{z}}$ onto the $\bs{z}$-space.

Denote by 
\be\label{appeq:Proj-For}
\widetilde{\bs{F}}(\bs{z}):=\int_{\Sigma} \bs{F}_{\bs{z}}(\bs{y}, \bs{z})\, \dd \pi(\bs{y}|\bs{z})
\ee
 the projected force defined using the conditional distribution 
 \be\label{appeq:cond-pdf}
 \dd \pi(\bs{y}|\bs{z}) = p_{\ds}(\bs{y},\bs{z})/\widetilde{p}_{\ds}(\bs{z})\,\dd\bs{y}.
 \ee
We can learn $\widetilde{\bs{F}}(\bs{z})$ via the following force projection loss
\begin{equation}\label{appeq:P-For-Loss}
\operatorname{L_{P-For}}(\widetilde{\bs{G}})=\int_{\Omega}\big|\bs{F}_{\bs{z}}(\bs{y}, \bs{z})- \widetilde{\bs{G}}(\bs{z})\big|^2\,  \dd \pi(\bs{y}, \bs{z}).
\end{equation}

To justify \eqref{appeq:P-For-Loss}, we note that 
\begin{align*}
& \int_{\Omega}\big|\bs{F}_{\bs{z}}(\bs{y}, \bs{z})- \widetilde{\bs{G}}(\bs{z})\big|^2\,  \dd \pi(\bs{y}, \bs{z})\\
= &  \int_{\Omega} \Big(|\bs{F}_{\bs{z}}(\bs{y}, \bs{z})|^2+ |\widetilde{\bs{G}}(\bs{z})|^2\Big)\,  \dd \pi(\bs{y}, \bs{z}) -2\int_{\Omega}\bs{F}_{\bs{z}}(\bs{y}, \bs{z})\cdot \widetilde{\bs{G}}(\bs{z})\,  \dd \pi(\bs{y}, \bs{z})\\
 =: & P_1-2P_2.
\end{align*}
 The term $P_2$ can be simplified as
\begin{align*}
P_2 & = \int_{\widetilde{\Omega}}\left[\int_{\Sigma}\bs{F}_{\bs{z}}(\bs{y},\bs{z})\dd \pi(\bs{y}|\bs{z})\right] \cdot \widetilde{\bs{G}}(\bs{z}) \, \widetilde{p}_{\ds}(\bs{z}) \,\dd \bs{z} = \int_{\widetilde{\Omega}} \widetilde{\bs{F}}(\bs{z}) \cdot \widetilde{\bs{G}}(\bs{z}) \, \widetilde{p}_{\ds}(\bs{z}) \,\dd \bs{z}.
\end{align*}
 Therefore, we have the equivalence
 \be
\min \operatorname{L_{P-For}}(\widetilde{\bs{G}})\quad \Longleftrightarrow\quad 
\min \operatorname{\widetilde{L}_{P-For}}(\widetilde{\bs{G}}),
\ee
where
\begin{equation*}
\operatorname{\widetilde{L}_{P-For}}(\widetilde{\bs{G}}):=\int_{\widetilde{\Omega}}\big|\widetilde{\bs{F}}(\bs{z})- \widetilde{\bs{G}}(\bs{z})\big|^2  \widetilde{p}_{\ds}(\bs{z}) \,\dd \bs{z}.
\end{equation*}
From the analysis above we can conclude that $\widetilde{F}(\bs{z})$ minimizes the loss in \eqref{appeq:P-For-Loss}. 

\subsection{HJB equation for the reduced potential}

In this subsection, we show that the reduced potential $\widetilde{U}$ satisfies the projected HJB equation
\begin{align}\label{appeq:P-HJB}
\widetilde{\bs{F}} & \cdot \nabla_{\bs{z}} \widetilde{U}  +|\nabla_{\bs{z}} \widetilde{U}|^2 -D \Delta_{\bs{z}} \widetilde{U}-D \nabla_{\bs{z}} \cdot \widetilde{\bs{F}} =0\,,
\end{align}
with asymptotic BC $\widetilde{U}\rightarrow \infty$ as $|\bs{z}|\rightarrow \infty$, or the reflecting BC $(\widetilde{\bs{F}}+\nabla_{\bs{z}} \widetilde{U}) \cdot \widetilde{\bs{n}}=0$ on $\partial \widetilde{\Omega}$, where  $\widetilde{\bs{n}}$ denotes the unit outer normal on $\partial\widetilde{\Omega}$.  We will only consider the rectangular domain case here. The argument for the unbounded case is similar. 

Recall that $p_{\ds}(\bs{x})$ satisfies the FPE  
\begin{equation}\label{appeq:fp}
\nabla \cdot(p_{\ds}\bs{F}) - D \Delta p_{\ds}=0.
\end{equation}
Integrating both sides of \eqref{appeq:fp} on $\Sigma$ with respect to $\bs{y}$ and utilizing the boundary condition $\bs{J}_{\ds}\cdot \bs{n}= 0$, where $\bs{J}_{\ds}=p_\ds\bs{F}-D\nabla p_\ds$, we get
\begin{equation}\label{appeq:fp-integrated}
\nabla_{\bs{z}} \cdot\Big(\int_{\Sigma}\bs{F}_{\bs{z}} p_{\ds}\,\dd \bs{y}\Big) - D \Delta_{\bs{z}} \widetilde{p}_{\ds}=0.
\end{equation}
Taking \eqref{appeq:Proj-For} and \eqref{appeq:cond-pdf} into account, we obtain
\begin{equation}\label{appeq:red-fp}
\nabla_{\bs{z}} \cdot\big(\widetilde{p}_{\ds}\widetilde{\bs{F}}\big) - D \Delta_{\bs{z}} \widetilde{p}_{\ds}=\nabla_{\bs{z}}\cdot \widetilde{\bs{J}}=0,
\end{equation}
i.e., a FPE for $\widetilde{p}_{\ds}(\bs{z})$ with the reduced force field $\widetilde{\bs{F}}$, where $\widetilde{\bs{J}}:=\widetilde{p}_{\ds}\widetilde{\bs{F}}  - D \nabla_{\bs{z}} \widetilde{p}_{\ds}$. The corresponding boundary condition can be also derived by integrating the original BC $\bs{J}_\ds\cdot \bs{n}  =0$
on $\Sigma$ with respect to $\bs{y}$  for $\bs{z}\in \partial\widetilde{\Omega}$, which gives 
\be\label{appeq:P-RBC}
\widetilde{\bs{J}}\cdot \widetilde{\bs{n}}=\big(\widetilde{p}_{\ds}\widetilde{\bs{F}} - D \nabla_{\bs{z}} \widetilde{p}_{\ds}\big)\cdot \widetilde{\bs{n}}=0.
\ee
Substituting the relation $\widetilde{p}_{\ds}(\bs{z})=\exp(-\widetilde{U}(\bs{z})/D)$ into \eqref{appeq:red-fp} and \eqref{appeq:P-RBC}, we get \eqref{appeq:P-HJB} and the corresponding reflecting BC after some algebraic manipulations.

\section{State-dependent diffusion coefficients}\label{Sec E: coefficients}

In this section, we study the EPR loss for NESS systems with a state-dependent diffusion coefficient.

Consider the It\^o SDE
\be\label{appeq:SDE-Var}
\frac{\dd \bs{x}(t)}{\dd t}=\bs{F}(\bs{x}(t)) + \sqrt{2D}{\sigma(\bs{x}(t))}\dot{\bs{w}}
\ee
with the state-dependent diffusion matrix $\sigma(\bs{x})$. Under the same assumptions as in MT Section 1.1, we have the FPE
\begin{equation}\label{appeq:fp-A}
\nabla \cdot(p_{\ds} \bs{F}) - D \nabla^2:\left(p_{\ds}a \right)=0.
\end{equation}
We show that the high dimensional landscape function $U$ of \eqref{appeq:SDE-Var} minimizes the EPR loss 
\begin{equation}\label{appeq:EPR-Var}
\operatorname{L_{V-EPR}}(V) = \int_{\Omega}\left|\bs{F}^v(\bs{x})+ a(\bs{x}) \nabla V(\bs{x})\right|_{a^{-1}(\bs{x})}^2 \dd \pi{(\bs{x})},
\end{equation}
where $\bs{F}^v(\bs{x}) := \bs{F}(\bs{x}) - D \nabla \cdot a(\bs{x})$ and $|\bs{u}|_{a^{-1}(\bs{x})}^2:=\bs{u}^\top a^{-1}(\bs{x})\bs{u}$ for $\bs{u}\in \mathbb{R}^d$. 

To justify \eqref{appeq:EPR-Var}, we first note that \eqref{appeq:fp-A} can be rewritten as
\be\label{appeq:FPE-V}
\nabla \cdot(p_{\ds}\bs{F}^v  - D a \nabla p_{\ds})=0\,,
\ee
which, together with the BC, implies the orthogonality relation
\begin{align}\label{appeq:FPE-V-weak}
\int_{\Omega}\big(\bs{F}^v+ a \nabla U\big)\cdot \nabla W\,\dd \pi = 0
\end{align}
for a suitable test function $W(\bs{x})$. Following the same reasoning used in establishing \eqref{appeq:Orth} and utilizing \eqref{appeq:FPE-V-weak}, we have 
\begin{align*}
& \int_{\Omega}\left|\bs{F}^v+ a  \nabla V\right|_{a^{-1}}^2 \,\dd \pi\\
= & \int_{\Omega}\big|\bs{F}^v+ a \nabla U
+ a  \nabla (V- U)\big|_{a^{-1}}^2\, \dd \pi\\
= & \int_{\Omega}|\bs{F}^v+ a  \nabla U\big|_{a^{-1}}^2 \,\dd \pi   + \int_{\Omega} \big|a \nabla (V- U)\big|_{a^{-1}}^2 \,\dd \pi.
\end{align*}
The last expression implies that $U(\bs{x})$ is the unique minimizer of $\operatorname{L_{V-EPR}}(V)$ up to a constant.

The above derivation for the state-dependent diffusion case will permit us to construct the landscape for the chemical Langevin dynamics, which will be studied in future works.

\newpage
\addcontentsline{toc}{section}{\bf Part 2: Computation}
\noindent{\color{blue} \textbf{PART 2: COMPUTATION}}

\vspace*{4mm}

Now we present the computational details omitted in the MT. We will first demonstrate the motivation for enhanced EPR, and then provide the training details and problem settings for landscape construction with original coordinates and dimension reduction. As mentioned in MT, we refer to the enhanced loss as Eq.~\eqref{eq:enh-loss-sm}.

\section{Motivation for enhanced EPR}\label{method: motivation for enhanced epr}

\begin{figure*}[!htpb]
    \centering
    \includegraphics[width=1.0\textwidth]{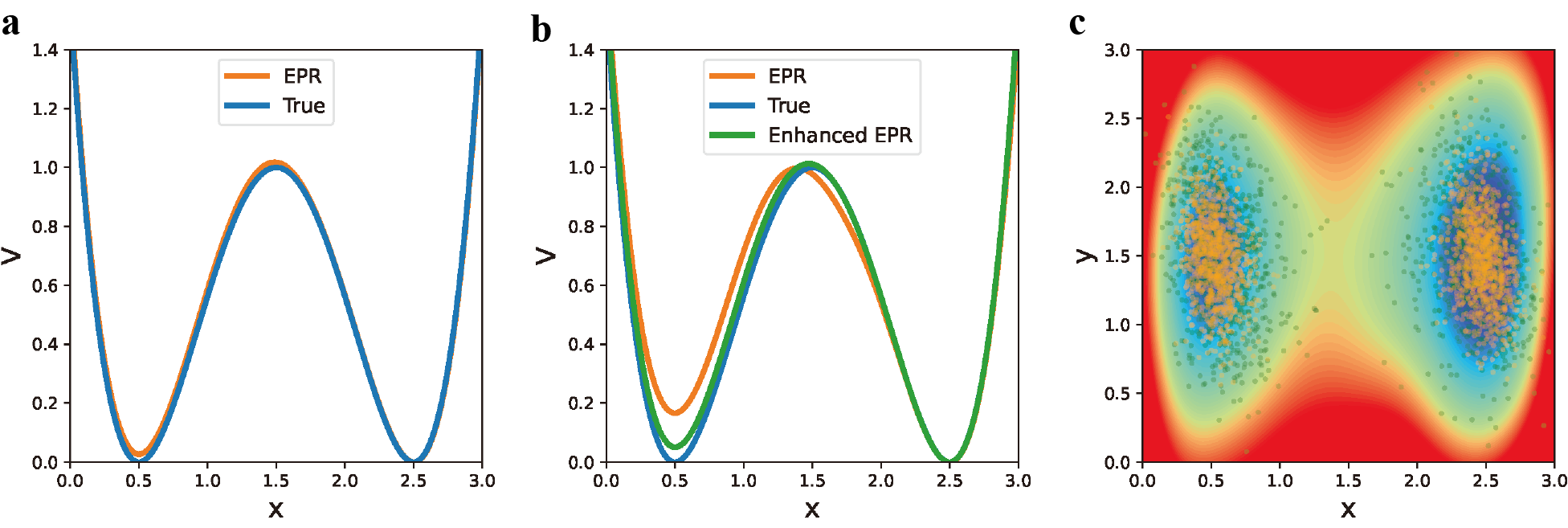}
    \caption{\textbf{An illustration for the motivation of enhanced EPR.} (a) and (b) show the comparisons of the learned potentials and true solution on the line $y=1.5$ in the toy model with $D=0.1$ and $D=0.05$, respectively. (c) shows the filled contour plot of the potential learned by only the EPR loss. The orange points are samples from the simulated invariant distribution with $D=0.05$, while green points are enhanced samples simulated from a more diffusive distribution with $D^\prime=0.1$, which are used in the enhanced EPR.}
    \label{fig:dw2d_fm_fh}
\end{figure*}

We utilize enhanced samples to achieve a better coverage of the transition domain. Note that the EPR loss equation (3) in MT requires training data sampled according to $\pi(\bs{x})$. The accuracy of the learned $V(\bs{x};\theta^*)$ (more precisely, the accuracy of the gradient $\nabla V(\bs{x};\theta^*)$) using equation (3) in MT is guaranteed only in the ``visible'' domain of $\pi$, i.e., regions that are covered by sample points. However, the enhanced EPR framework~\eqref{eq:enh-loss-sm} combines both the EPR loss and the HJB, which allows the use of sample data that better covers the domain $\Omega$ (e.g., more samples in the transition regions between meta-stable states and near the boundaries of the visible domain). This approach is particularly compelling when the diffusion coefficient D is relatively small. 

The single EPR loss works well for the toy model with a relatively large diffusion coefficient $D=0.1$. A slice plot of the potential at $y=1.5$ in Fig.~\textbf{\ref{fig:dw2d_fm_fh}a} shows that the learned solution with EPR loss coincides well with the analytical solution. The relative root mean square error (rRMSE) and the relative mean absolute error (rMAE) have the mean and the standard deviation of $0.084\pm0.006$ and $0.066\pm0.008$ over 3 runs, respectively. 

However, when decreasing $D$ to $0.05$, the samples from the simulated invariant distribution mainly stay in the double wells (orange points in Fig.~\textbf{\ref{fig:dw2d_fm_fh}c}) and are away from the transition region between the two wells. For this reason, as shown in Fig.~\textbf{\ref{fig:dw2d_fm_fh}b}, the result with a single EPR loss captures the profile of the two wells, but these two parts of the profile are not accurately connected in the transition region where there are few samples, making the left well a bit higher than the right one. We then generate enhanced samples $(\bs{x}'_i)_{1\le i\le N'}$ using $D'=0.1$, which have better coverage of the transition region (green points in Fig.~\textbf{\ref{fig:dw2d_fm_fh}a}).  Incorporating these enhanced samples, we learn the potential with the enhanced EPR loss and the result agrees better with the true solution (Fig.~\textbf{\ref{fig:dw2d_fm_fh}b}).

Though the enhanced EPR is more recommended in general situations, we remark that the single EPR loss can achieve competitive performance as long as the samples cover the domain effectively. This numerical experience works for all of the considered models.

\section{Landscape with original coordinates} \label{Sec F: 2d models}

In this section, we will describe the additional setup where we utilize enhanced EPR for landscape construction with original coordinates in MT. This section includes a 2D toy model, a 2D biological system with a limit cycle~\cite{Wang08}, a 2D multi-stable system~\cite{Wang11} and a 12D Gaussian mixture model (GMM). 

\subsection{Training details and additional results for benchmark problems}\label{Sec F.1.}

We begin by simulating the SDE using the Euler-Maruyama scheme with reflecting boundary conditions until time $\mathsf{T}=1000$, starting from $10000$ different initial states. This simulation yields $N=10000$ final states, which are used as the training dataset $(\mathbf{x}_i)_{1\le i\le N}$ to approximate the invariant distribution for EPR. For the 2D toy problem and 2D multi-stable problem, we employ a time-step size of 0.1, whereas for the 2D limit cycle problem, we use a smaller time-step size of 0.01. We train the network with a batch size of 1024 and a learning rate of 0.001 using the Adam~\cite{kingma2015adam} optimizer for 3000 epochs in the toy models, 5000 epochs in the multi-stable model and 10000 epochs in the limit cycle model. When focusing on single EPR loss in the toy model with $D = 0.05$, we extend to 8000 epochs to ensure adequate learning. At each training epoch, we update the dataset by performing one step Euler-Maruyama scheme to make it closer to the invariant distribution.  For the comparison with normalizing flows, we train a neural spline flow~\cite{NSF19}.\footnote{https://github.com/643VincentStimper/normalizing-flows} We repeat 4 blocks of the rational quadratic spline with three layers of 64 hidden units and a followed LU linear permutation. The flow model is trained by the Adam optimizer with the learning rate 0.0001 for 20000 epochs, based on the same SDE-simulated dataset as the enhanced EPR. To ensure the robustness of our results, we perform our experiments across 5 random seeds, which helps to account for variability and establish the reliability of our findings.

\begin{table*}[]
\centering
\caption{Evaluation on various $\mu(\bs x)$ and $\hat{\lambda}_1$ for HJB alone and enhanced EPR over 5 random seeds. }
\label{tab:add_results}
\begin{tabular}{c|c||cccc|cccc}
\toprule
   \multirow{3}{*}{\rotatebox{90}{\textbf{ }}}                                 &  \textbf{Metrics}      & \multicolumn{4}{c|}{\textbf{rRMSE}} & \multicolumn{4}{c}{\textbf{rMAE}} \\ \cmidrule{2-10}
  &
  \multirow{2}{*}{\diagbox{$\mu(\bs x)$}{$\hat{\lambda}_1$}} &
  \textbf{HJB} &
  \multicolumn{3}{c|}{\textbf{Enhanced EPR}} &
  \textbf{HJB} &
  \multicolumn{3}{c}{\textbf{Enhanced EPR}} \\
                                      &        & $\times$ 0.0     & $\times$ 0.1    & $\times$ 1.0    & $\times$ 10.0    & $\times$ 0.0    & $\times$ 0.1    & $\times$ 1.0    & $\times$ 10.0   \\ \midrule
\multirow{7}{*}{\rotatebox{90}{\textbf{Toy, $D=0.1$}}}   & $D^\prime=1D$  &     \makecell{\small{0.034} \\ \myscriptsize{$\pm$0.016}}     &  ------      &   ------     &    ------     & \makecell{\small{0.029} \\ \myscriptsize{$\pm$0.014}}        &  ------       &   ------      &   ------      \\
                                      & $D^\prime=2D$  &  \makecell{\small{0.168} \\ \myscriptsize{$\pm$0.010}}        &  \makecell{\small{0.062} \\ \myscriptsize{$\pm$0.033}}      &    \makecell{\textbf{\small{0.028}} \\ $\pm$\myscriptsize{0.010}}     &     \makecell{\small{0.038} \\ $\pm$\myscriptsize{0.015}}     &   \makecell{\small{0.068} \\ \myscriptsize{$\pm$0.009}}        &  \makecell{\small{0.108} \\ \myscriptsize{$\pm$0.054}}      &   \makecell{\textbf{\small{0.026}} \\ \myscriptsize{$\pm$0.010}}   &  \makecell{\small{0.032} \\ \myscriptsize{$\pm$0.013}}       \\
                                      & $D^\prime=5D$  &   \makecell{\small{0.225} \\ $\pm$\myscriptsize{0.071}}       &     \makecell{\small{0.138} \\ $\pm$\myscriptsize{0.021}}    &   \makecell{\small{0.080} \\ $\pm$\myscriptsize{0.011}}      &    \makecell{\small{0.046} \\ $\pm$\myscriptsize{0.018}}      &    \makecell{\small{0.123} \\ $\pm$\myscriptsize{0.097}}     & \makecell{\small{0.032} \\ \myscriptsize{$\pm$0.025}}        &  \makecell{\small{0.030} \\ \myscriptsize{$\pm$0.011}}      &  \makecell{\small{0.030} \\ \myscriptsize{$\pm$0.019}}      \\
                                      & $D^\prime=10D$ &    \makecell{\small{0.258} \\ $\pm$\myscriptsize{0.078}}     &    \makecell{\small{0.182} \\ $\pm$\myscriptsize{0.033}}     &   \makecell{\small{0.095} \\ $\pm$\myscriptsize{0.010}}      &   \makecell{\small{0.067} \\ $\pm$\myscriptsize{0.012}}       &    \makecell{\small{0.167} \\ $\pm$\myscriptsize{0.110}}      &  \makecell{\small{0.062} \\ \myscriptsize{$\pm$0.040}}      &   \makecell{\small{0.038} \\ \myscriptsize{$\pm$0.010}}     &   \makecell{\small{0.037} \\ \myscriptsize{$\pm$0.009}}     \\ \midrule
\multirow{7}{*}{\rotatebox{90}{\textbf{Toy, $D=0.05$}}}  & $D^\prime=1D$  &     \makecell{\small{0.191} \\ \myscriptsize{$\pm$0.218}}     &   ------      &   ------     &   ------      &  \makecell{\small{0.160} \\ $\pm$\myscriptsize{0.179}}       &    ------     &     ------    &    ------     \\
                                      & $D^\prime=2D$  &    \makecell{\small{0.267} \\ \myscriptsize{$\pm$0.116}}       &   \makecell{\small{0.058} \\ \myscriptsize{$\pm$0.045}}     &    \makecell{\textbf{\small{0.054}} \\ \myscriptsize{$\pm$0.020}}    &     \makecell{\small{0.111} \\ \myscriptsize{$\pm$0.083}}    &  \makecell{\small{0.164} \\ $\pm$\myscriptsize{0.120}}       &  \makecell{{\small{0.049}} \\ \myscriptsize{$\pm$0.036}}      &   \makecell{\textbf{\small{0.048}} \\ \myscriptsize{$\pm$0.019}}     &  \makecell{{\small{0.096}} \\ \myscriptsize{$\pm$0.069}}      \\
                                      & $D^\prime=5D$  &   \makecell{\small{0.661} \\ \myscriptsize{$\pm$0.126}}        &   \makecell{\small{0.417} \\ \myscriptsize{$\pm$0.162}}     &   \makecell{\small{0.204} \\ \myscriptsize{$\pm$0.213}}     &     \makecell{\small{0.055} \\ \myscriptsize{$\pm$0.028}}    &    \makecell{\small{0.555} \\ $\pm$\myscriptsize{0.093}}     &    \makecell{{\small{0.328}} \\ \myscriptsize{$\pm$0.155}}    &  \makecell{{\small{0.127}} \\ \myscriptsize{$\pm$0.184}}      &  \makecell{{\small{0.029}} \\ \myscriptsize{$\pm$0.008}}      \\
                                      & $D^\prime=10D$ &   \makecell{\small{0.544} \\ \myscriptsize{$\pm$0.061}}        &   \makecell{\small{0.485} \\ \myscriptsize{$\pm$0.126}}     &  \makecell{\small{0.227} \\ \myscriptsize{$\pm$0.132}}      &   \makecell{\small{0.191} \\ \myscriptsize{$\pm$0.190}}      &  \makecell{\small{0.489} \\ $\pm$\myscriptsize{0.057}}       &  \makecell{{\small{0.410}} \\ \myscriptsize{$\pm$0.096}}      &  \makecell{{\small{0.157}} \\ \myscriptsize{$\pm$0.136}}      &  \makecell{{\small{0.122}} \\ \myscriptsize{$\pm$0.168}}      \\ \midrule
\multirow{7}{*}{\rotatebox{90}{\textbf{Multi-stable}}}   & $D^\prime=1D$  &     \makecell{\small{0.255} \\ \myscriptsize{$\pm$0.007}}     &   ------      &   ------      &    ------      &  \makecell{\small{0.241} \\ \myscriptsize{$\pm$0.004}}       &  ------      &  ------      &  ------      \\
                                      & $D^\prime=2D$  &   \makecell{\small{0.249} \\ \myscriptsize{$\pm$0.015}}       &    \makecell{\small{0.069} \\ \myscriptsize{$\pm$0.045}}      &   \makecell{\small{0.110} \\ \myscriptsize{$\pm$0.028}}       &  \makecell{\small{0.146} \\ \myscriptsize{$\pm$0.044}}         &    \makecell{\small{0.228} \\ \myscriptsize{$\pm$0.011}}       &   \makecell{\small{0.065} \\ \myscriptsize{$\pm$0.046}}        &    \makecell{\small{0.106} \\ \myscriptsize{$\pm$0.030}}      &   \makecell{\small{0.142} \\ \myscriptsize{$\pm$0.047}}       \\
                                      & $D^\prime=5D$  &   \makecell{\small{0.628} \\ \myscriptsize{$\pm$0.046}}       &   \makecell{\small{0.061} \\ \myscriptsize{$\pm$0.031}}      &   \makecell{\small{0.090} \\ \myscriptsize{$\pm$0.042}}      &      \makecell{\small{0.128} \\ \myscriptsize{$\pm$0.043}}    &    \makecell{\small{0.553} \\ \myscriptsize{$\pm$0.063}}      &   \makecell{\small{0.059} \\ \myscriptsize{$\pm$0.032}}       &   \makecell{\small{0.088} \\ \myscriptsize{$\pm$0.046}}       &   \makecell{\small{0.124} \\ \myscriptsize{$\pm$0.044}}       \\
                                      & $D^\prime=10D$ &  \makecell{\small{0.630} \\ \myscriptsize{$\pm$0.051}}        &  \makecell{\small{0.307} \\ \myscriptsize{$\pm$0.036}}       &  \makecell{\textbf{\small{0.067}} \\ \myscriptsize{$\pm$0.047}}       &  \makecell{\small{0.118} \\ \myscriptsize{$\pm$0.032}}        &     \makecell{\small{0.550} \\ \myscriptsize{$\pm$0.066}}     &     \makecell{\small{0.197} \\ \myscriptsize{$\pm$0.036}}     &   \makecell{\textbf{\small{0.066}} \\ \myscriptsize{$\pm$0.049}}       &   \makecell{\small{0.118} \\ \myscriptsize{$\pm$0.034}}       \\ \midrule
\multirow{7}{*}{\rotatebox{90}{\textbf{Limit Cycle}}} & $\sigma=0.0$ &  \makecell{\small{0.231} \\ \myscriptsize{$\pm$0.048}}        &   ------     &     ------   &    ------     &  \makecell{\small{0.140} \\ \myscriptsize{$\pm$0.021}}        &     ------     &    ------      &  ------        \\
                                      & $\sigma=0.05$ &   \makecell{\small{0.287} \\ \myscriptsize{$\pm$0.175}}       &   \makecell{\small{0.256} \\ \myscriptsize{$\pm$0.181}}      &   \makecell{\textbf{\small{0.070}} \\ \myscriptsize{$\pm$0.016}}      &    \makecell{\small{0.136} \\ \myscriptsize{$\pm$0.002}}      &  \makecell{\small{0.165} \\ \myscriptsize{$\pm$0.108}}       &   \makecell{\small{0.149} \\ \myscriptsize{$\pm$0.107}}      &   \makecell{\textbf{\small{0.063}} \\ \myscriptsize{$\pm$0.020}}      &  \makecell{\small{0.108} \\ \myscriptsize{$\pm$0.003}}       \\
                                      & $\sigma=0.1$  &  \makecell{\small{0.503} \\ \myscriptsize{$\pm$0.133}}        &   \makecell{\small{0.372} \\ \myscriptsize{$\pm$0.173}}      & \makecell{\small{0.373} \\ \myscriptsize{$\pm$0.181}}        &    \makecell{\small{0.122} \\ \myscriptsize{$\pm$0.008}}      &  \makecell{\small{0.313} \\ \myscriptsize{$\pm$0.092}}       & \makecell{\small{0.222} \\ \myscriptsize{$\pm$0.119}}        &   \makecell{\small{0.200} \\ \myscriptsize{$\pm$0.090}}      &  \makecell{\small{0.101} \\ \myscriptsize{$\pm$0.006}}       \\
                                      & $\sigma=0.2$ &  \makecell{\small{0.578} \\ \myscriptsize{$\pm$0.010}}        &   \makecell{\small{0.556} \\ \myscriptsize{$\pm$0.013}}      &   \makecell{\small{0.509} \\ \myscriptsize{$\pm$0.025}}      &     \makecell{\small{0.298} \\ \myscriptsize{$\pm$0.055}}     &  \makecell{\small{0.383} \\ \myscriptsize{$\pm$0.025}}       &   \makecell{\small{0.355} \\ \myscriptsize{$\pm$0.030}}      &  \makecell{\small{0.303} \\ \myscriptsize{$\pm$0.040}}       &   \makecell{\small{0.175} \\ \myscriptsize{$\pm$0.029}}      \\ \bottomrule
\end{tabular}
\end{table*}

For a fair comparison, we fix $\lambda_2$ at 1.0 across all models. The parameters we use in the MT is $\lambda_1=1.0$ for toy model and multi-stable model, and $\lambda_1=0.01$ for the limit cycle model. As discussed in the MT, the selection of $\lambda_1$ aims to balance the two terms of the loss function. Based on our experience, systems with higher entropy production rates—typically featuring more non-gradient components in their force decomposition—require a smaller $\lambda_1$, as in the limit cycle problem. However, the specific choice of $\lambda_1$ is relatively robust. As demonstrated in Table~\ref{tab:add_results}, we scrutinize the stability of $\lambda_1$ by testing its sensitivity within the enhanced EPR method. We adjust $\lambda_1$ by multiplying it by a set of factors \texttt{\{0.0, 0.1, 1.0, 10.0\}} as $\hat{\lambda}_1$ for different experiments. Then $\hat{\lambda}_1=0.0$ corresponds to the scenario of using HJB alone. Despite the varying distributions of $\mu (\mathbf{x})$, our selected parameters typically yield satisfactory outcomes. Even with a relatively small $\hat{\lambda}_1$, enhanced EPR outperforms solving by HJB alone. The performance enhancements are more pronounced when relatively large values of $\hat{\lambda}_1$ are used.

We also evaluate different distributions of samples using the HJB loss, \textit{i.e.}, $\mu (\bs x)$. For enhanced EPR, using invariant distribution in the HJB loss term is ineffective, as HJB loss is involved to cover the transition domain. Consequently, there is no need to evaluate $\mu (\bs x)$ with $D'=1D$ or $\sigma=0$ for enhanced EPR. In the case of the HJB loss, our numerical analysis suggests that overly diffusive samples can lead to significantly worse outcomes, underscoring the importance of capturing the critical domains within the loss. As shown in Table~\ref{tab:add_results}, the optimal results for HJB alone are achieved using the invariant distribution or a distribution slightly more extensive than that. However, with the same enhanced samples from $\mu (\bs x)$, introducing a non-zero $\lambda_1$ enhances performance compared to HJB alone, particularly when the sample domain is so diffusive that solving HJB alone performs poorly. This emphasizes the robustness of enhanced EPR to variations in $\mu (\bs x)$ and its advantage over solving HJB alone. Finally, we emphasize the optimal outcome for $\hat{\lambda}_1 = 1.0 \times \lambda_1$ across various $\mu(\bs x)$ for each problem by bolding it in Table~\ref{tab:add_results}, serving as the result for enhanced EPR in Table 1 of the MT. While this may not be the best result among all the combinations of $\mu(\bs x)$ and $\hat{\lambda}_1$, it is a commonly robust choice that can be readily achieved without dedicated tuning.

\subsection{2D toy model}\label{methods: 2d toy model} 

To verify the applicability and accuracy of our method, we initially apply it to a toy model with the driving force 
\be\label{eq:Test1}
\bs{F}(\bs{x})=-(I+A) \nabla U_0(\bs{x}),
\ee
where $A\in \mathbb{R}^{d\times d}$ is a constant skew-symmetric matrix, i.e., $A^\top=-A$, and $U_0$ is some known function. With this choice of $\bs{F}$, one can check that the true potential landscape is simply $U(\bs{x})=U_0(\bs{x})$. In particular, the system becomes reversible when $A=0$. We construct a 2D toy model with the double-well potential as
\be
U_0(\bs{x}) =  ((x - 1.5)^2 - 1.0)^2 + 0.5(y - 1.5)^2,
\ee
where $\bs{x}=(x,y)^\top$. We take the anti-symmetric matrix
\be
A = \begin{bmatrix}
	 0 & 0.5\\
	-0.5 & 0
	 \end{bmatrix},
\ee
which introduces a counter-clockwise rotation for a focusing central force field. This sets up a simple non-equilibrium system. In this model, we have the force decomposition 
$$
\bs{F}(\bs{x})= -\nabla U_0(\bs{x})+ \bs{l}(\bs{x}),\ \bs{l}(\bs{x})=-A \nabla U_0(\bs{x})
$$
and 
$$\bs{l}(\bs{x})\cdot \nabla U_0(\bs{x}) = 0, \quad \nabla \cdot \bs{l}(\bs{x})=0 $$
hold in the pointwise sense. Therefore, the identity
$\bs{l}(\bs{x})\cdot \nabla U_0(\bs{x}) + D\nabla \cdot \bs{l}(\bs{x})=0$ is satisfied for any $D>0$ and, following the discussions in MT, we have constructed a non-reversible system with analytically known double-well potential which can be used to verify the 
accuracy of the learned potential. We focus on the domain $\Omega=[0, 3]\times[0, 3]$. To fix the extra shifting degree of freedom of the potential function, we set the minimum of the potential to be zero and only plot the result on the domain $\{\bs{x} | V(\bs{x}) \leq 30D\}$ in Fig.~3 in MT since it is the domain of practical interest.

\subsection{Two-dimensional limit cycle model}\label{sec:F.2}

We apply our approach to the limit cycle dynamics with a Mexican-hat shape landscape~\cite{Wang08}. 

Before introducing the concrete model, let us make the following observation. For any SDE like
\be\label{appeq:SDE}
\frac{\dd \bs{x}}{\dd t}=\bs{F}(\bs{x}) + \sqrt{2D}\dot{\bs{w}},
\ee
the corresponding steady FPE is
\begin{equation*}
\nabla \cdot(p_{\ds}\bs{F}) - D \Delta p_{\ds}=0.
\end{equation*}
If we make the transformation 
$$\bs{F}\rightarrow \kappa \bs{F},\  D\rightarrow \kappa D$$
in \eqref{appeq:SDE}, then the steady state PDF 
$$p_{\ds}(\bs{x}) = \exp(-\frac{U(\bs{x})}{D})=\exp(-\frac{\kappa U(\bs{x})}{\kappa D})$$ 
will not change. The transformation only changes the timescale of the dynamics \eqref{appeq:SDE} from $t_0$ to $\kappa t_0$. However, the potential changes from $U$ to $\kappa U$ if we utilize the drift $\kappa \bs{F}(\bs{x})$ and noise strength $\kappa D$ in the system \eqref{appeq:SDE}. This observation allows us  to set the scale of $U$ to be $O(1)$ by adjusting $\kappa$ suitably for a specific problem. An alternative approach to accomplish this task is by choosing $\bs{F}$ to be $\kappa \bs{F}$ in the EPR loss.

We take $D=0.1$ and consider the stochastic dynamics \eqref{appeq:SDE} with $\bs{F} = (F_x, F_y)$ and
\begin{align}
F_x(x, y) & =\kappa\left(\frac{\alpha^2+x^2}{1+x^2} \frac{1}{1+y}-a x\right), \\
F_y(x, y) & =\frac{\kappa}{\tau_0}\left(b-\frac{y}{1+c x^2}\right),
\end{align}
where the parameters are $\kappa=100, \alpha=a=b=0.1, c=100$, and $\tau_0=5$. Here the choice of $\kappa=100$ is made such that $U\sim O(1)$ following \cite{Lin22}. We focus on the domain $\Omega=[0, 8]\times[0, 6]$. As explained in the paragraph above, our setting corresponds to the case $D=0.1/\kappa=0.001$ for the force field considered in \cite{Wang08}.

\subsection{Two-dimensional multi-stable model}
\label{sec:F.4}

We also study the dynamics \eqref{appeq:SDE} for a multi-stable system~\cite{Wang11} with $\bs{F} = (F_x, F_y)$ and
\begin{align}
F_x(x, y) & =\frac{a x^n}{S^n+x^n}+\frac{b S^n}{S^n+y^n}-k_1 x, \\
F_y(x, y) & =\frac{a y^n}{S^n+y^n}+\frac{b S^n}{S^n+x^n}-k_2 y,
\end{align}
where the parameters are $a = b = k_1=k_2=1$, $S=0.5$, and $n=4$. We focus on the domain $\Omega=[0, 3]\times[0, 3]$ and present the results for $D=0.01$ in MT.

\subsection{Twelve-dimensional GMM model: high-dimensional potential}\label{sec: F.5}
As mentioned in MT Methods 3.4, we can construct models with known potential with driving force like Eq.~\eqref{eq:Test1}.
The true solution $U_0$ is denoted by $U_0(\bs{x}) = -D \log p_0(\bs{x})$, where $p_0$ is the probability distribution of the GMM, as defined in MT Section 1.3. We randomly generate a matrix $A_0$ with elements uniformly in $[0, 1]$, then obtain an skew-symmetric matrix by $A = (A_0 - A_0^\top)/2$. The non-reversible dynamics is then constructed with driving force in \eqref{eq:Test1}. We set the noise strength as $D=0.01$ for this problem and use $D^\prime=5D$ for enhanced samples. We use a three-layer neural network with 80 hidden states per layer. The data size is 10000 and we use Adam with a learning rate of 0.001 and batch size of 2048. We train enhanced EPR with $\lambda_1 = 0.1, \lambda_2 = 1.0$ for 5000 epochs.

\section{Landscape with reduced coordinates}\label{Sec H: high models}

In this section, we provide additional details and results related to the dimension reduction problem. First, we present additional experiments of the 3D Ferrell's cell cycle problem~\cite{Ferrell11}. We then demonstrate an 8D limit cycle dynamics \cite{Wang10}, in which interesting properties emerge with projected force $\widetilde{\bs{G}}(\bs{z}; \theta^*)$ and potential $\widetilde{V}(\bs{z};\theta^*)$. Finally, we present a 52D multi-stable dynamics \cite{Li13}, on which we compare the reduced potential $\widetilde{V}(\bs{z};\theta^*)$ by two dimensionality reduction approaches.

\subsection{Ferrell's three-ODE model: reduced potential}\label{sec: G.1}
\begin{figure}[!h]
    \centering
     \includegraphics[width=0.75\textwidth]{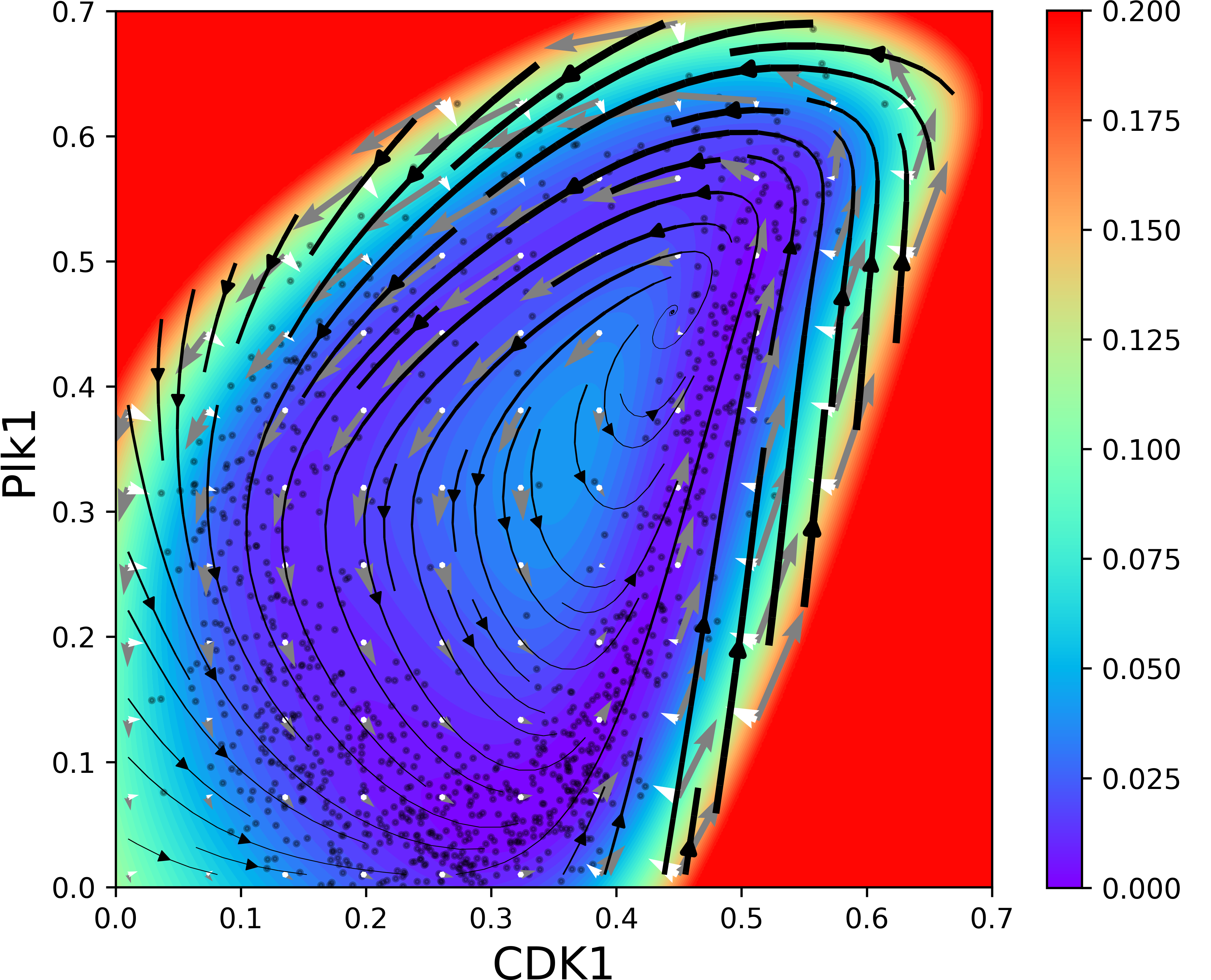}
    \caption{Streamlines of the projected force $\widetilde{\bs{G}}(\bs{z})$ and filled contour plot of the reduced potential $\widetilde{V}(\bs{z};\theta^*)$ for Ferrell's three-ODE model learned by enhanced EPR. The projected force field $\widetilde{\bs{G}}(\bs{z})$ is decomposed into the gradient part $-\nabla \widetilde{V}(\bs{z};\theta^*)$ (white arrows) and the non-gradient part (gray arrows). The length of an arrow corresponds to the magnitude of the vector. The solid dots are samples from the simulated invariant distribution.}
    \label{fig:bio3d_cycle}
\end{figure}
We utilize the dimensionality reduction method on Ferrell's three-ODE model for a simplified cell cycle dynamics \cite{Ferrell11}, where 
the concentrations of the cyclin-dependent protein kinase ($\CDK 1$), Polo-like kinase 1 ($\Plk 1$), and the anaphase-promoting complex ($\APC$), denoted by 
$$
x=[\CDK 1],\ y=[\Plk 1],\ z=[\APC]
$$
respectively, obey the ODEs in the domain $\Omega=[0, 1]^3$
\begin{align}
F_x(x, y, z) & =\alpha_1-\beta_1 x \frac{z^{n_1}}{K_1^{n_1}+z^{n_1}},\\
F_y(x, y, z) & =\alpha_2\left(1-y\right) \frac{x^{n_2}}{K_2^{n_2}+x^{n_2}}-\beta_2 y,\\
F_z(x, y, z) & =\alpha_3\left(1-z\right) \frac{y ^{n_3}}{K_3^{n_3}+y^{n_3}}-\beta_3 z,
\end{align}
with $\alpha_1=0.1, \alpha_2=\alpha_3=\beta_1=3, \beta_2=\beta_3=1, K_1=K_2=K_3=0.5, n_1=n_2=8,$ and $n_3=8$. Since our focus in this paper is on the methodology of constructing the potential landscape, we refer the interested readers to the literature~\cite{Ferrell11} for concrete biological meaning of the considered variables. We add the noise scale $D=0.01$ with isotropic temporal Gaussian white noise.  For this particular problem, we employ three-layer neural networks with $80$ hidden units in each layer. We generate enhanced samples $(\mathbf{x}^\prime_i)_{1\leq i\leq 10000}$ by simulating from a more diffusive distribution with $D'=5D$. The projected force $\widetilde{\mathbf{G}}(\mathbf{z};\theta^*)$ is trained by the loss~\eqref{appeq:P-For-Loss} for $1000$ epochs using Adam optimizer with a learning rate 0.001 and a batch size 2048. The reduced variables $\mathbf{z}=(x,y)^\top$ are utilized during this training phase. Subsequently, we train the projected potential $\widetilde{V}(\mathbf{z};\theta^*)$ for $4000$ epochs using the enhanced EPR loss defined in \eqref{eq:enh-loss-sm}, with the chosen values of $\lambda_1=0.1$ and $\lambda_2=1.0$. As shown in Fig.~\ref{fig:bio3d_cycle}, the obtained reduced potential shows a plateau in the centering region and a local-well tube domain along the reduced limit cycle.

\subsection{Eight-dimensional complex system: reduced potential}\label{sec: G.2}

\begin{figure}[!h]
    \centering
    \includegraphics[width=0.75\textwidth]{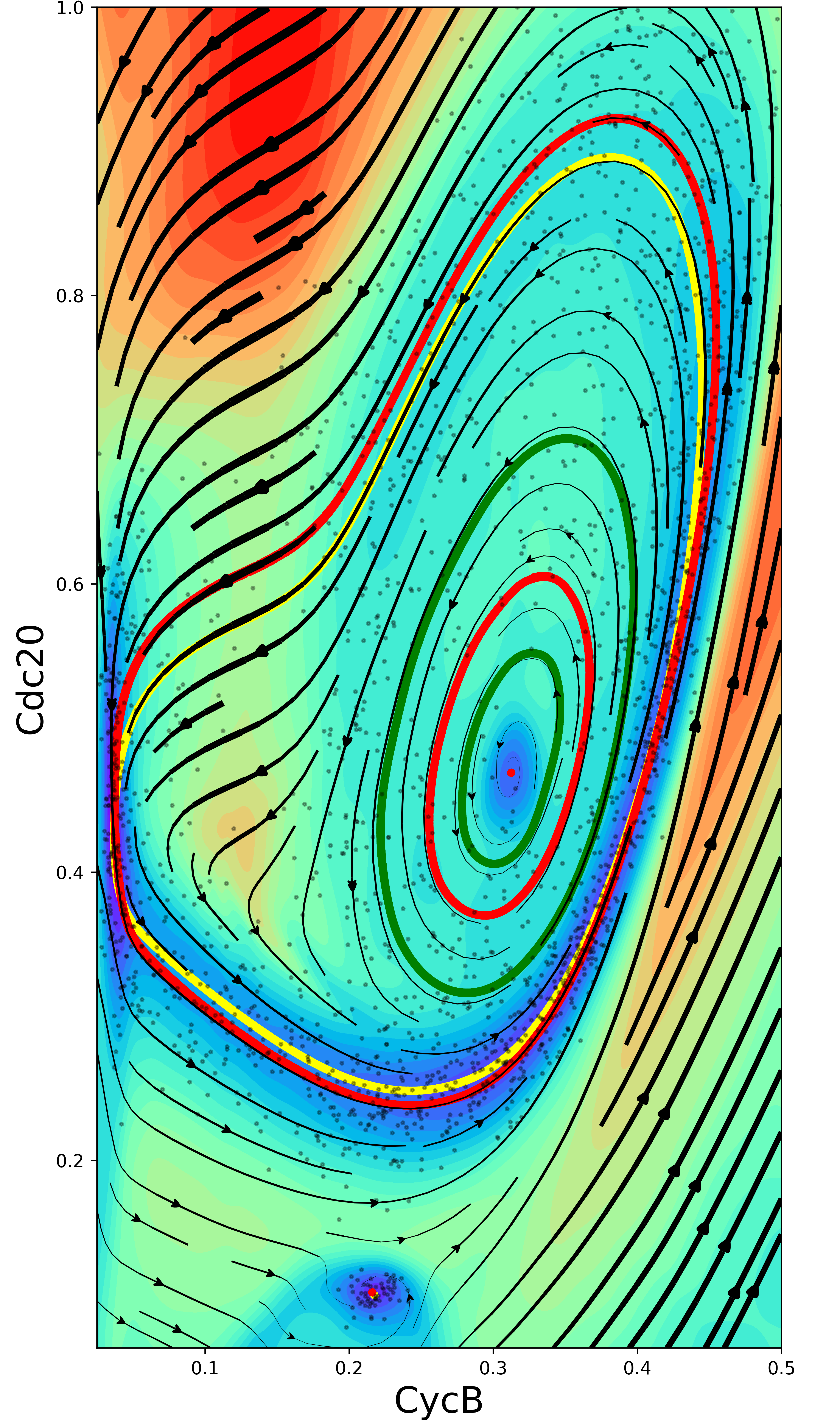}
    \caption{Streamlines and limit sets of the projected force field of the 8D cell cycle model by two reduced variables CycB and Cdc20. The outer red circle is the stable limit cycle of the reduced force field corresponding to the yellow circle as the projection of the original high-dimensional limit cycle. The inner red circle, red dot and two green circles are stable and unstable limit sets of the reduced dynamics, respectively, which are virtual in high dimensions.}
    \label{fig:bio8d_ode}
\end{figure}

We consider an 8D system in which the dynamics and parameters are the same as in the supplementary information of~\cite{Wang10}. We take CycB and Cyc20 as the reduction variable $\bs{z}$, and set the mass in this problem as $m=0.8$.

We first point out that the noise strength $D=0.0005$ used in \cite{Wang10} is not suitable here in training neural networks since this would lead to a potential of order $O(10^{-5})$. Using the idea in SI Section~\ref{sec:F.2}, we amplify the original force field $\bs{F}$ in \cite{Wang10} by $\kappa=1000$ times, and take $D=0.01$ for the transformed force field. This amounts to setting $D=10^{-5}$ for the original force field, which is even smaller than the parameter considered in \cite{Wang10}. We simulate the SDE without boundaries with timestep $10^{-5}$ until $\mathsf{T}=5$, starting from initial states drawn from a uniform distribution in $[0, 1.25]^8$. The enhanced samples $(\bs{x}'_i)_{1\le i\le N'}$ are obtained by adding Gaussian perturbations with a standard deviation $\sigma=0.05$ to the SDE-simulated dataset $(\bs{x}_i)_{1\le i\le N}$. We only keep the data within the biologically meaningful domain of $[0, 1.25]^8$ of size 25000 for computation.

We use three-layer networks with 80 hidden states in each layer for both force and potential. For the projected force, we train 2000 epochs to obtain $\widetilde{\bs{G}}(\bs{z};\theta^*)$. Then we conduct the enhanced EPR \eqref{eq:enh-loss-sm} with $\lambda_1=0.01, \lambda_2=1.0$ for 10000 epochs. We use Adam with a learning rate of 0.001 and batch size of 8192.

In Fig.~\ref{fig:bio8d_ode}, we present a more detailed picture of the reduced dynamics for the 8D model than Fig. 5c in MT. Specifically, we further show two unstable limit cycles of the projected force field obtained by reversed time integration (two green circles in Fig.~\ref{fig:bio8d_ode}). These two unstable limit cycles play the role of separatrices between the neighboring stable limit sets. This picture occurs due to the fact that the landscape of the considered system is very flat in the centering region. These inner limit sets are virtual in high dimensions, but they naturally appear in the reduced dynamics on the plane. Similar features might also occur in other reduced dynamics in two dimensions.

\subsection{Fifty-two-dimensional multi-stable system:  high-dimensional and reduced potentials}\label{sec: G.3}

For fairness, we all use a neural network with three-layer and 20 hidden states in each layer to denote the reduced force $\widetilde{\bs{G}}(\bs{z};\theta^*)$ and potential $\widetilde{V_1}(\bs{z};\theta^*)$, and 80 hidden states for the 52D potential $V(\bs{x}; \theta^*)$. We use enhanced samples simulated from a more diffusive distribution with $D^\prime = 2D$. The data size is 20000 and we use Adam with a learning rate of 0.001 and batch size of 2048, still. We use $\lambda_1=10.0, \lambda_2=1.0$ in enhanced EPR \eqref{eq:enh-loss-sm}. We train the force $\widetilde{\bs{G}}(\bs{z};\theta^*)$ for 1000 epochs and conduct $\widetilde{V_1}(\bs{z};\theta^*)$ by enhanced EPR with for 5000 epochs. We also learn a 52D potential $V(\bs{x}; \theta^*)$ by enhanced EPR for 5000 epochs and then project it to $\widetilde{V_2}(\bs{z};\theta^*)$ by~\eqref{appeq:GP-Loss} for 1000 epochs.

\end{document}